\documentclass[twocolumn]{aastex701}

\newcommand{\db}{\textsc{Dense-Basis}}
\newcommand{\bp}{\textsc{Bagpipes}}
\newcommand{\zphot}{$z_\mathrm{phot}$}
\newcommand{\zspec}{$z_\mathrm{spec}$}
\newcommand{\logM}{log$(M_\star/M_\odot)$}

\begin{document}

\title{Massive Galaxies Form Early and Gray:\\ Stellar Assembly and Dust Attenuation at $\mathbf{z > 3.5}$ from CAPERS}

\author[0000-0003-4922-0613]{Katherine Chworowsky}
\affiliation{Department of Astronomy, The University of Texas at Austin, Austin, TX 78712 USA}
\affiliation{Cosmic Frontier Center, The University of Texas at Austin, Austin, TX, USA}
\email{k.chworowsky@utexas.edu}

\author[0000-0001-8519-1130]{Steven L. Finkelstein}
\affiliation{Department of Astronomy, The University of Texas at Austin, Austin, TX 78712 USA}
\affiliation{Cosmic Frontier Center, The University of Texas at Austin, Austin, TX, USA}
\email{stevenf@astro.as.utexas.edu}

\author[0000-0003-1282-7454]{Anthony J. Taylor}
\affiliation{Department of Astronomy, The University of Texas at Austin, Austin, TX 78712 USA}
\affiliation{Cosmic Frontier Center, The University of Texas at Austin, Austin, TX, USA}
\email{anthony.taylor@austin.utexas.edu }

\author[0000-0003-4965-0402]{Alexa M.\ Morales}\altaffiliation{NSF Graduate Research Fellow}
\affiliation{Department of Astronomy, The University of Texas at Austin, Austin, TX 78712 USA}
\affiliation{Cosmic Frontier Center, The University of Texas at Austin, Austin, TX, USA}
\email{alexa.morales@utexas.edu}

\author[0000-0001-5414-5131]{Mark Dickinson}
\affiliation{NSF's National Optical-Infrared Astronomy Research Laboratory, 950 N. Cherry Ave., Tucson, AZ 85719, USA}
\email{mark.dickinson@noirlab.edu}

\author[0000-0003-3466-035X]{{L. Y. Aaron} {Yung}}
\altaffiliation{Giacconi Postdoctoral Fellow}
\affiliation{Space Telescope Science Institute, 3700 San Martin Dr., Baltimore, MD 21218, USA}
\email{yung@stsci.edu}

\author[0000-0002-7959-8783]{Pablo Arrabal Haro}
\affiliation{Center for Space Sciences and Technology, UMBC, 5523 Research Park Dr, Baltimore, MD 21228 USA }
\affiliation{Astrophysics Science Division, NASA Goddard Space Flight Center, 8800 Greenbelt Rd, Greenbelt, MD 20771, USA}
\email{pablo.arrabalharo@nasa.gov}

\author[0000-0001-8534-7502]{Bren E. Backhaus}
\affil{Department of Physics and Astronomy, University of Kansas, Lawrence, KS 66045, USA}
\email{bren.backhaus@ku.edu}

\author[0000-0001-8863-2472]{Davide Bevacqua}
\affil{Physics and Astronomy Department, Tufts University, 574 Boston Avenue, Medford, MA 02155, USA}
\affiliation{CINAF—Osservatorio Astronomico di Brera, via Brera 28, 20121 Milano, Italy}
\email{davide.bevacqua@inaf.it}

\author[0000-0003-2332-5505]{\'{O}scar Ch\'{a}vez Ortiz}
\affiliation{Department of Astronomy, The University of Texas at Austin, Austin, TX 78712 USA}
\affiliation{Cosmic Frontier Center, The University of Texas at Austin, Austin, TX, USA}
\email{chavezoscar009@utexas.edu}

\author[0000-0002-1482-5818]{Adam C. Carnall}
\affiliation{Institute for Astronomy, University of Edinburgh, Royal Observatory, Edinburgh, EH9 3HJ, UK}
\email{adamc@roe.ac.uk}

\author[0000-0002-7622-0208]{Callum T. Donnan}
\affiliation{NSF's National Optical-Infrared Astronomy Research Laboratory, 950 N. Cherry Ave., Tucson, AZ 85719, USA}
\email{callum.donnan@noirlab.edu}

\author[0000-0002-7831-8751]{Mauro Giavalisco}
\affiliation{University of Massachusetts Amherst, 710 North Pleasant Street, Amherst, MA 01003-9305, USA}
\email{mauro@umass.edu}

\author[0000-0002-3301-3321]{Michaela Hirschmann}
\affiliation{Institute of Physics, Laboratory for Galaxy Evolution, EPFL, Observatoire de Sauverny, Chemin Pegasi 51, CH-1290 Versoix, Switzerland}
\email{michaela.hirschmann@epfl.ch}

\author[0000-0001-9298-3523]{Kartheik G. Iyer}
\affiliation{Center for Computational Astrophysics, Flatiron Institute, 162 Fifth Avenue, New York, NY 10010, USA}
\email{kartheikiyer@gmail.com}

\author[0000-0002-6610-2048]{Anton M. Koekemoer}
\affiliation{Space Telescope Science Institute, 3700 San Martin Drive,
Baltimore, MD 21218, USA}
\email{koekemoer@stsci.edu}

\author[0000-0003-2366-8858]{Rebecca L.\ Larson}
\altaffiliation{Giacconi Postdoctoral Fellow}
\affil{Space Telescope Science Institute, 3700 San Martin Drive, Baltimore, MD 21218, USA}
\email{rlarson@stsci.edu}

\author[0000-0003-1581-7825]{Ray A. Lucas}
\affiliation{Space Telescope Science Institute, 3700 San Martin Drive, Baltimore, MD 21218, USA}
\email{lucas@stsci.edu}

\author[0000-0002-6149-8178]{Jed McKinney}
\altaffiliation{NASA Hubble Fellow}
\affiliation{Department of Astronomy, The University of Texas at Austin, Austin, TX 78712 USA}
\affiliation{Cosmic Frontier Center, The University of Texas at Austin, Austin, TX, USA}
\email{jed.mckinney@austin.utexas.edu }

\author[0000-0003-4368-3326]{Derek J. McLeod}
\affiliation{Institute for Astronomy, University of Edinburgh, Royal Observatory, Edinburgh EH9 3HJ, UK}
\email{derek.mcleod@ed.ac.uk}

\author[0000-0001-7503-8482]{Casey Papovich}
\affiliation{Department of Physics and Astronomy, Texas A\&M University, College Station, TX, 77843-4242 USA}
\affiliation{George P.\ and Cynthia Woods Mitchell Institute for Fundamental Physics and Astronomy, Texas A\&M University, College Station, TX, 77843-4242 USA}
\email{papovich@tamu.edu}

\author[0000-0003-4528-5639]{Pablo G. P\'erez-Gonz\'alez}
\affiliation{Centro de Astrobiolog\'{\i}a (CAB), CSIC-INTA, Ctra. de Ajalvir km 4, Torrej\'on de Ardoz, E-28850, Madrid, Spain}
\email{pgperez@cab.inta.csic.es}

\author[0000-0001-9495-7759]{Lu Shen}
\affiliation{Department of Physics and Astronomy, Texas A\&M University, College Station, TX, 77843-4242 USA}
\affiliation{George P.\ and Cynthia Woods Mitchell Institute for
 Fundamental Physics and Astronomy, Texas A\&M University, College Station, TX, 77843-4242 USA}
\email{lushen@tamu.edu}

\author[0000-0002-6748-6821]{Rachel S. Somerville}
\affiliation{Center for Computational Astrophysics, Flatiron Institute, 162 Fifth Avenue, New York, NY 10010, USA}
\email{rsomerville@flatironinstitute.org }

\author[0000-0002-2906-2200]{Laura Sommovigo}
\affiliation{Center for Computational Astrophysics, Flatiron Institute, 162 Fifth Avenue, New York, NY 10010, USA}
\email{lsommovigo@flatironinstitute.org}

\author[0000-0002-0827-9769]{Thomas M. Stanton}
\affiliation{Institute for Astronomy, University of Edinburgh, Royal Observatory, Edinburgh EH9 3HJ, UK}
\email{t.stanton@ed.ac.uk}

\author[0000-0002-9373-3865]{Xin Wang}
\affiliation{School of Astronomy and Space Science, University of Chinese Academy of Sciences (UCAS), Beijing 100049, China}
\affiliation{National Astronomical Observatories, Chinese Academy of Sciences, Beijing 100101, China}
\affiliation{Institute for Frontiers in Astronomy and Astrophysics, Beijing Normal University, Beijing 102206, China}
\email{xwang@ucas.ac.cn}

\author[0000-0003-3903-6935]{Stephen M. Wilkins}
\email{S.Wilkins@sussex.ac.uk}
\affiliation{Astronomy Centre, University of Sussex, Falmer, Brighton BN1 9QH, UK}

\author[0000-0002-7051-1100]{Jorge A. Zavala}
\email{jzavala@umass.edu}
\affiliation{University of Massachusetts Amherst, 710 North Pleasant Street, Amherst, MA 01003-9305, USA} 

\collaboration{all}{and The CAPERS collaboration}

\begin{abstract}
The stellar mass assembly of massive galaxies in the first few billion years of cosmic history remains a central challenge in galaxy formation. Galaxies with $M_\star \gtrsim 10^{10}M_\odot$ observed at $z \gtrsim 4$ must grow rapidly under conditions of intense gas accretion, feedback, and dust production. Observationally, their star-formation histories (SFHs) have been poorly constrained due to degeneracies inherent to broadband photometry. The advent of \textit{JWST} enables direct spectroscopic access to detailed continuum shapes and rest-frame optical diagnostics at high redshift, providing a critical opportunity to reconstruct formation timescales of massive early galaxies.
Here, we investigate massive galaxies using joint spectro-photometric SED fitting of \textit{JWST}/NIRSpec prism spectroscopy from the CANDELS-Area Prism Epoch of Reionization Survey (CAPERS). Our sample comprises 148 galaxies selected photometrically with \logM $> 9.5$ at $z > 3.5$. 
We find that the most massive galaxies (\logM $> 10.5$) preferentially exhibit shallow, gray dust attenuation curves, consistent with higher dust optical depths and large grain sizes. We also find significant diversity in the time at which galaxies form 25\% of their stellar mass. While formation timescales converge toward later cosmic times, galaxies with lower sSFR ($\lesssim -9$) at the observation epoch formed significantly earlier than systems with higher sSFRs. Across the full mass range, inferred assembly times are systematically earlier than model predictions, suggesting more rapid early growth than currently captured theoretically. These results underscore the importance of spectroscopic constraints and flexible SFH and dust models for reconstructing high-redshift massive galaxy formation histories.

\end{abstract}

\keywords{\uat{Galaxies}{573}}


\section{Introduction}  \label{sec:intro}

Massive galaxies at high redshift provide a unique window into the physics of galaxy formation during the first few billion years of cosmic history. Building stellar masses in excess of $10^{10}\,M_\odot$ at $z \gtrsim 4$ requires sustained or highly efficient star formation under conditions where gas accretion, feedback, and dynamical timescales differ substantially from those in the local Universe. As a result, the star-formation histories (SFHs) of massive galaxies encode critical information about how star formation is regulated, and when galaxies transition between active growth and suppression of star-formation.
Historically, observational studies of high-redshift massive galaxies have focused primarily on integrated properties such as stellar mass, luminosity, and star-formation rate, derived largely from broadband photometry \citep[e.g.][]{Muzzin2013, Duncan2014, Tomczak2014, Song2016, Weaver2022, PerezGonzalez2008}. While these measurements have established the existence and abundance of massive systems up to a redshift of $z \sim 4$, they provide limited insight into the timescales over which these galaxies assembled their stellar mass. In particular, photometry alone struggles to distinguish between galaxies that formed the bulk of their mass rapidly at early epochs and those that experienced more extended or episodic growth, due to degeneracies between age, dust attenuation, metallicity, and nebular emission \citep{Conroy2013, Papovich2001, Finkelstein2010, Narayanan2024}.

The launch of \textit{JWST} \citep{Gardner2006, Gardner2023} has opened a new observational regime for studying massive galaxies at high redshift. With its sensitivity and wavelength coverage extending into the rest-frame optical at $z>4$, \textit{JWST} enables direct constraints on stellar continua, Balmer and 4000~\AA\ breaks, and strong nebular emission lines through spectro-photometric observations. These data substantially improve our ability to reconstruct SFHs, particularly when combined with flexible, non-parametric modeling approaches that do not impose strong assumptions about the shape or smoothness of star formation over time \citep{Iyer2019, Carnall2019SFH, Leja2019}.

At the same time, interpretation of spectro-photometric datasets critically depends on an accurate treatment of dust attenuation. Dust not only reshapes the observed spectral energy distribution by reddening stellar continua and attenuating nebular emission, but also introduces degeneracies that directly impact inferred stellar ages, star-formation rates, and mass-weighted formation times \citep[e.g.][]{Calzetti2000, Conroy2013, Pacifici2023, Leja2019}. While attenuation curves derived from local star-forming galaxies have been widely adopted in high-redshift studies, a growing body of evidence indicates that dust attenuation laws vary significantly as a function of galaxy mass, metallicity, and optical depth, and may evolve with redshift \citep{Sommovigo2025, CharlotFall2000, Salim2018, Kriek2013, Narayanan2026}. Observations at intermediate redshifts reveal systematic trends in both attenuation-curve slope and UV bump strength with galaxy properties, suggesting that dust geometry and grain composition are closely linked to the physical conditions of the interstellar medium \citep[e.g.][]{GarnBest2010, Salim2018, Reddy2015, Shivaei2020}. Recent \textit{JWST} observations have begun to extend these constraints to $z \gtrsim 6$, finding that complex dust populations can already be in place at early cosmic times \citep[e.g.][]{Witstok2023, Fisher2025, Markov2025, Ormerod2025}. These results indicate that flexible attenuation prescriptions are essential for accurately modeling dust attenuation, which is necessary to obtain reliable constraints on the stellar mass formation histories of massive galaxies in the early Universe.

Since launch, \textit{JWST} has enabled the first large samples of $z\gtrsim4$ galaxies with rest-frame optical coverage, and several studies have begun to infer their SFHs. Analyses using deep \textit{JWST}/NIRCam imaging have applied flexible or non-parametric SED-fitting methods to constrain stellar ages and formation times for galaxies at high redshifts \citep[e.g.][]{Tacchella2023, Santini2025, Lisiecki2025}. While \textit{JWST} has shown that massive galaxies with evolved stellar populations exist within the first $\sim1$–2 Gyr of cosmic time, the constraints on detailed SFH shape and early mass assembly remain limited by degeneracies between age, dust, metallicity, and nebular emission that persist when relying on broadband photometry alone. Spectroscopic programs with \textit{JWST}/NIRSpec have provided more direct constraints on recent star formation and dust attenuation through rest-frame optical continua and emission lines \citep[e.g.][]{Carnall2023, Bunker2023, Curtis-Lake2023, PerezGonzalez2025}. However, many early NIRSpec studies target relatively small samples, limiting their ability to reconstruct detailed SFHs for statistically meaningful samples of galaxies. As a result, the formation histories of galaxies that have already assembled large amounts of stellar mass at high redshift remain weakly constrained.

The CANDELS Area Prism Epoch Reionization Survey (CAPERS) builds on this progress by combining deep \textit{JWST}/NIRSpec prism spectroscopy with homogeneous multi-band \textit{JWST}/NIRCam photometry over the legacy CANDELS extragalactic field. The depth of the prism data enables constraints on both rest-frame optical continua and strong nebular features, while the survey design yields a substantial sample ($N>100$) of massive galaxies at high redshifts. 
In this work, we investigate the formation histories of a sample of massive ($\log M_\star \gtrsim 9.5$) galaxies at $z \gtrsim 3.5$ using joint spectro-photometric SED fitting of \textit{JWST}/NIRSpec prism spectroscopy from CAPERS, together with multi-band \textit{HST}/ACS + \textit{JWST}/NIRCam photometry. By focusing on galaxies that have already assembled substantial stellar mass at early cosmic times we can begin to constrain the onset of star-formation, and confront them with modern simulation predictions to understand the physical processes governing it.

The structure of this paper is as follows. We describe the \textit{JWST} spectroscopic and multi-wavelength photometric data used in this analysis, along with the procedure adopted to place the spectra and photometry on a consistent absolute flux scale in Section~\ref{sec:data}. Section~\ref{sec:sample} details the sample selection and the spectro-photometric SED-fitting methodology, including the assumptions and priors used to model star-formation histories and other physical parameters. In Section~\ref{sec:analysis}, we present the inferred galaxy properties and the derived star-formation history metrics used to characterize stellar mass assembly. We discuss the implications of these results for the growth of massive galaxies and dust attenuation at high redshifts in Section~\ref{sec:discussion}, and summarize our main conclusions in Section~\ref{sec:summary}. Throughout this paper, we assume a cosmology of H$_0$=70 km s$^{-1}$ Mpc$^{-1}$, $\Omega_M$=0.3, $\Omega_\Lambda$ = 0.7.  We assume a \citet{Chabrier2003} IMF (converting literature and fit values to a Chabrier IMF when necessary), and all magnitudes given are in the AB system \citep{ABmag}.

\section{Data} \label{sec:data}

\subsection{Spectroscopy} \label{subsec:CAPERS}
CAPERS (GO-6368, PI: M.~Dickinson) is a \textit{JWST} Cycle~3 large Treasury Program designed to investigate galaxy evolution during the first 1.5~Gyr of cosmic history. CAPERS uses the NIRSpec micro-shutter assembly (MSA; \citealt{Ferruit2022}) in PRISM mode ($R \sim 100$) to obtain spectroscopy of galaxies selected from deep, multiband \textit{HST} and \textit{JWST}/NIRCam imaging in three legacy CANDELS fields \citep{Grogin2011,Koekemoer2011}. The observed sources are selected from the COSMOS \citep{Scoville2007}, Ultra-Deep Survey (UDS; \citealt{Lawrence2007}), and Extended Groth Strip (EGS; \citealt{Davis2007}) fields. Both COSMOS and UDS were imaged with NIRCam as part of the Public Release Imaging for Extragalactic Research survey (PRIMER; GO-1837, Dunlop et al., \textit{in prep.}), while EGS was observed through the Cosmic Evolution Early Release Science program (CEERS; GO-1345, \citealt{Finkelstein2025}).

CAPERS targets seven MSA pointings in each of the PRIMER-UDS, PRIMER-COSMOS, and CEERS-EGS fields. In each pointing, three independent MSA configurations are executed. 
Each MSA configuration employs a three-shutter slitlet with nodding between the three shutters. The effective exposure time per MSA configuration is 5,690~s, resulting in a total effective exposure time of 17,069~s per pointing for high priority targets observed in all three configurations. 
For this work, we use data from all three CAPERS fields observed between December 2024 and January 2026. This is about 94\% of the complete CAPERS data set, missing only one MSA configuration still to be observed in COSMOS and three in UDS.
The targets to be observed by CAPERS are selected based on photometric redshifts from the photometric catalogs (\S~\ref{subsec:photometry}) and assigned a baseline MSA weight following $\log_{10}(w_0) = 0.5\times z_\mathrm{phot}$ for $0 < z_\mathrm{phot} < 10$ and $\log_{10}(w_0) = 5$ for $z_\mathrm{phot} > 10$. Targets belonging to science subsamples contributed by astronomers  within and outside the CAPERS collaboration receive an additional priority boost above this baseline (see more in \S~\ref{subsec:spec_select}. Objects that are very faint (magnitude $m > 28$) in the three reddest NIRCam broad bands F277W, F356W and F444W were downweighted, but this is not the case for any of the massive galaxy candidates considered here.

The NIRSpec data were reduced using the JWST Calibration Pipeline\footnote{\url{https://github.com/spacetelescope/jwst}} \citep{JWSTcalib} version~1.20.2 and Calibration Reference Data System (CRDS) context \texttt{pmap\_1464}, one-dimensional spectra were extracted using an optimal extraction technique following \citet{Horne1986}. The full reduction incorporates a number of customized reduction steps, described by \citet{ArrabalHaro2023}. 
The spectroscopic redshifts used for SED-fitting in this work were determined using a combination of automated tools and visual inspection. First, each spectrum was fit with variety of software including \bp{} \citep{Carnall2019SFH}, \textsc{Cigale} \citep{CIGALE}, \textsc{msaexp} \citep{msaexp}, and custom tools. The resulting redshift solutions were then compared, and the best solution for each spectrum was selected via visual inspection. Additional details of the CAPERS observations, target selection, and data reduction will be presented in forthcoming papers from the collaboration.

\subsection{Photometric Catalogs} \label{subsec:photometry}
We use photometric catalogs from the Uniform Near-Infrared CatalOgs from Robust imagiNg project (UNICORN; S. Finkelstein et al. in prep), which include all {\it HST}/ACS and {\it JWST}/NIRCam data in our targeted fields. For the PRIMER fields these are from \textit{HST}: F435W, F606W, and F814W; from \textit{JWST}: F090W, F115W, F140M, F150W, F182M, F200W, F210M, F250M, F277W, F335M, F356W, F410M, F444W, and F480M (the medium bands beyond F410M come from the MINERVA survey (\citealt{Muzzin2013}), and cover half of the UDS field in the reductions used). For EGS, from \textit{HST}: F435W, F606W, and F814W; from \textit{JWST}: F090W, F115W, F150W, F200W, F277W, F356W, F410M, and F444W. The image reduction procedures followed the CEERS-team procedures detailed in the Appendix of \citet{Finkelstein2025}.  The catalogs are broadly based on the photometric procedures of \citet{Finkelstein2024}, thus we refer the reader there for detailed information. In brief, these catalogs are Source Extractor \citep{Bertin2010} based, using F277W+F356W as the detection image, and provide robust estimates of object colors, total fluxes, and photometric uncertainties. Photometric redshifts are including, using \textsc{Lazy} \footnote{\url{https://github.com/hollisakins/Lazy.jl}}, which is a Julia-based version of the template-based fitting code EaZY \citep{EAZY}.

\subsection{Spectra Rescaling} \label{subsec:spectra_norm}

The absolute flux calibration of the NIRSpec prism spectra is affected by wavelength-dependent slit losses and aperture mismatches between the spectroscopic extraction and the broadband photometry. To recover integrated galaxy properties that are consistent with the total photometric flux, we apply a slitloss rescaling that places each spectrum on the same absolute flux scale as the multi-band photometric catalog prior to spectro-photometric SED fitting. 

For each galaxy, we compute synthetic broadband fluxes by integrating the observed spectrum through the transmission curves of the \emph{JWST}/NIRCam filters used in the photometric catalog. Only filters in which both the synthetic spectroscopic flux and the observed photometric flux are detected at $\mathrm{S/N} \geq 2$ are used to derive the correction. The ratio of the observed photometric flux to the synthetic spectroscopic flux is fit linearly with a low-order Chebyshev polynomial (of degree $N_{\text{valid filters}} - 1$, with maximum degree $= 3$), with weights derived from the combined photometric and spectroscopic uncertainties. To avoid unphysical behavior outside the wavelength range spanned by the photometric constraints, the correction function is held fixed at its edge values beyond the bluest and reddest valid filters and is subsequently smoothed with a boxcar kernel with width 10 pixels.
Both the spectral flux and its associated uncertainty are multiplied by the resulting correction function to match the integrated broadband fluxes. 
While \bp\ has a built in \texttt{calib} method which performs a similar calibration process, we chose to perform the normalization to the data itself (prior to fitting) to avoid covariances between this correction and inference on physical properties. In \S~\ref{subsec:slitloss} we explore how this affects the recovered physical properties for the galaxies.

\section{Sample Selection} \label{sec:sample}

In order to infer the physical properties of our sample of massive high redshift galaxies, we combine deep multi-wavelength \textit{HST} and \textit{JWST} imaging with \textit{JWST}/NIRSpec prism spectroscopy, and use spectro-photometric SED-fitting techniques that account for flexible star-formation histories, dust attenuation, and nebular emission. Below, we outline the photometric SED fitting used to estimate redshifts and initial physical parameters and describe the selection criteria defining our final sample.

\begin{figure*}[th]
    \centering
    \includegraphics[width=\textwidth]{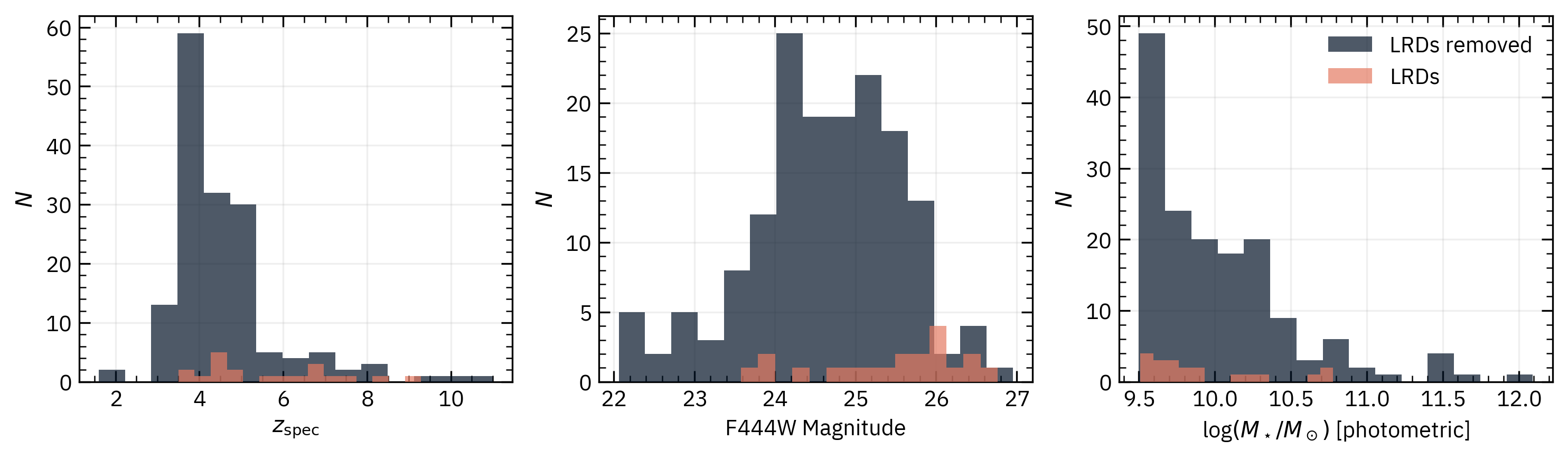}
    \caption{The distribution of our sample in vetted spectroscopic redshift, F444W magnitude, and the estimated stellar masses based on photometry (assuming no AGN component). The dark gray shaded region is the final sample of 148 galaxies which satisfy a photometric redshift \zphot$>3.5$ and photometrically inferred stellar mass of \logM $>9.5$ with LRDs removed. The red shaded region are the 23 LRDs removed based on the criteria described in \S~\ref{subsec:spec_select}.}
    \label{fig:sample_dist}
\end{figure*}

\subsection{Photometric SED fitting}\label{subsec:sed_fitting_phot}

To have photometrically determined stellar masses and redshifts for a parent sample, we perform SED fitting using \db\  \citep{Iyer2019} on the complete photometric catalogs from the EGS, PRIMER-COSMOS, and PRIMER-UDS fields. This allows us to select a parent sample and compare between fits from photometric catalogs from which sources are selected for follow-up, and how the estimated physical parameters vary when additional data is acquired.
Our SED-fitting with \db\ uses a prior determined from the \textsc{Lazy} redshift probability distribution, $P(z)$(\S~\ref{subsec:photometry}). As \db\ takes redshift priors in the form of a top-hat prior, to define redshift priors for each source, we identify the primary peak of the un-smoothed $P(z)$. To reduce the impact of small-scale fluctuations, the distribution is then smoothed using a two-stage moving-average filter. The location of the dominant peak is refined by searching for the maximum of the smoothed distribution within a limited window around the original peak position, while restricting any shift to be small in redshift space.
Redshift bounds are subsequently determined by tracing the smoothed $P(z)$ away from the peak until the distribution transitions from a steeply declining slope to a flatter profile on both the low- and high-redshift sides. This slope-based criterion adapts the width of the redshift interval to the intrinsic shape of the probability distribution, naturally producing narrow bounds for sharply peaked solutions and broader bounds for more extended or multimodal distributions.

We adopt photometric redshift priors derived from \textsc{Lazy} rather than fitting a completely free redshift simultaneously with the physical parameters in \db, motivated by the fact that template-based photometric redshift codes have been shown to recover accurate and robust redshifts across a wide range of galaxy populations \citep{EAZY}. By performing the SED fitting with a well-constrained photometric redshift prior from \textsc{Lazy}, we assume accurate redshift estimates and reduce degeneracies between redshift and stellar population properties that can be introduced in more flexible SED modeling codes.

We chose to perform catalog-level SED fitting with \db\ as it is capable of modeling large catalogs with non-parametric SFHs in a relatively short timeframe (order of $\sim$ms per source) by precomputing model grids given input priors and photometric broadbands.
\db\ uses a flexible non-parametric SFH represented by a Gaussian Process Model \citep[GPM; ][]{Iyer2019}, and stellar templates generated from \texttt{FSPS} \citep{Conroy2009, ConroyGunn2009} including implementation of nebular emission lines using \texttt{CLOUDY} \citep{CLOUDY, Byler2017}.
For this work, we define three ``shape'' parameters that describe the SFH: $\tau_{25}, \tau_{50},$ and $\tau_{75}$ (requiring the recovered SFH of the galaxy to form ``$x$'' percent of its total mass by time $\tau_x$). The catalogs were fit with \db\ assuming a \citet{Calzetti2000} dust law and a Chabrier initial mass function \citep[IMF;][]{Chabrier2003}. We impose a uniform (flat) prior in log space on the specific star formation rate (sSFR): log (sSFR/$ \textrm{yr}^{-1}) \in [-14, -7]$, an exponential prior on the dust attenuation over a wide range of values ($A_V \in [0, 4]$), and a uniform (in log-space) prior on the metallicity (log $Z/Z_\odot \in [-2, 0.3]$), with $Z_\odot = 0.02$ \citep[][]{Anders1989}. 

\subsection{Spectroscopic Selection}\label{subsec:spec_select}

To comprise our massive galaxy sample for further analysis,
we cross matched all galaxies observed spectroscopically by CAPERS with those in the photometric catalogs with photometric redshifts of \zphot\ $>3.5$ and \logM$>9.5$ (\S~\ref{subsec:sed_fitting_phot}), finding a sample of 176 sources. 
The spectra of all sources were visually inspected to assess data quality.
Three sources were removed due to them being oversubtracted in the employed nodded reduction due to the presence of nearby companions that enter the same slitlet in some of the nods (IDs: CAPERS-UDS-10997, CAPERS-COSMOS-35764, and CAPERS-COSMOS-21379). 

One possible contaminant in the selection for massive galaxies are ``little red dots" \citep[LRDs;][]{Matthee2024, Labbe2023Natur.616..266L, Chworowsky2024AJ....168..113C}.  As the rest-optical emission in these sources is likely due to accretion around a super-massive black hole, we aim to identify and remove them from our sample. We fit power-law continua to the rest-frame ultraviolet and optical portions of each spectrum, masking strong emission lines and regions affected by the Balmer and 4000\AA\ breaks. 
We identify LRDs following the definition of \citet{Akins2025}: We select sources exhibiting a negative UV spectral slope (UV slope $<0$) and a positive optical spectral slope (optical slope $>0$) in $f_\lambda$. And further require these sources to be compact in F444W, quantified using the compactness metric:
\begin{equation}
    C_{\mathrm{F444W}} = f_{\mathrm{F444W}}(d = 0.2^") / f_{\mathrm{F444W}}(d = 0.5^").
\end{equation}
Point sources are those with $C_{\mathrm{F444W}} > 0.5$, or those where at least 50\% of the F444W broadband emission within a $0.5"$ is contained within a $0.2"$ diameter aperture.
Galaxies satisfying both the spectral slope and point source criteria are classified as LRDs. 
We find that 20 objects satisfy this selection, implying that there is a significant AGN contribution in the rest-optical, for which stellar masses cannot be reliably estimated without explicitly modeling the AGN light, which is beyond the scope of this paper. 
We further remove 3 sources (IDs: CAPERS-EGS-19799, CAPERS-EGS-7761, and CAPERS-UDS-15322) which did not satisfy our LRD compactness criteria, but were identified as LRDs by \citet{Barro2025}. After applying these quality and classification cuts we find a total of 23 LRDs and exclude them from this analysis. Additionally, we identify two sources with significant broad $H\alpha$ emission, likely indicative of AGN activity (IDs: CAPERS-EGS-27615 and CAPERS-EGS-52661) and remove those from the analysis sample as well. 

Our final sample is comprised of 148 galaxies. The distributions of redshifts, magnitudes, and stellar masses for the final sample, as well as for the removed LRDs, are shown in Figure~\ref{fig:sample_dist}. As described in \S~\ref{subsec:CAPERS}, CAPERS spectroscopic prioritization gave higher weight to targets with higher photometric redshifts. Priorities were boosted further for contributed target subsamples of special interest. Of our final sample, 77 (52\%) carry no subsample flags in CAPERS and were observed at the default, redshift-dependent baseline priorities. The remaining 71 (48\%) are flagged for one or more contributed subsamples. The dominant flag in our final sample is for candidate massive galaxies with photometrically derived stellar masses of \logM$>10$ (26 galaxies, 18\%). Their priorities were boosted in order to better sample the most massive (and least numerous) galaxies at high redshift. 5 more objects belong to other subsamples that are potentially associated with massive systems, of which 2 objects were selected as candidate quiescent galaxies based on their photometric SEDs. Therefore, while nearly half of the final sample had targeting priority boosted with respect to the baseline, only 2 out of 148 (1.35\%) were prioritized as potentially quiescent massive galaxies. This massive galaxy spectroscopic sample should therefore have little bias with respect to past star formation histories.

\begin{figure}[th]
    \centering
    \includegraphics[width=\columnwidth]{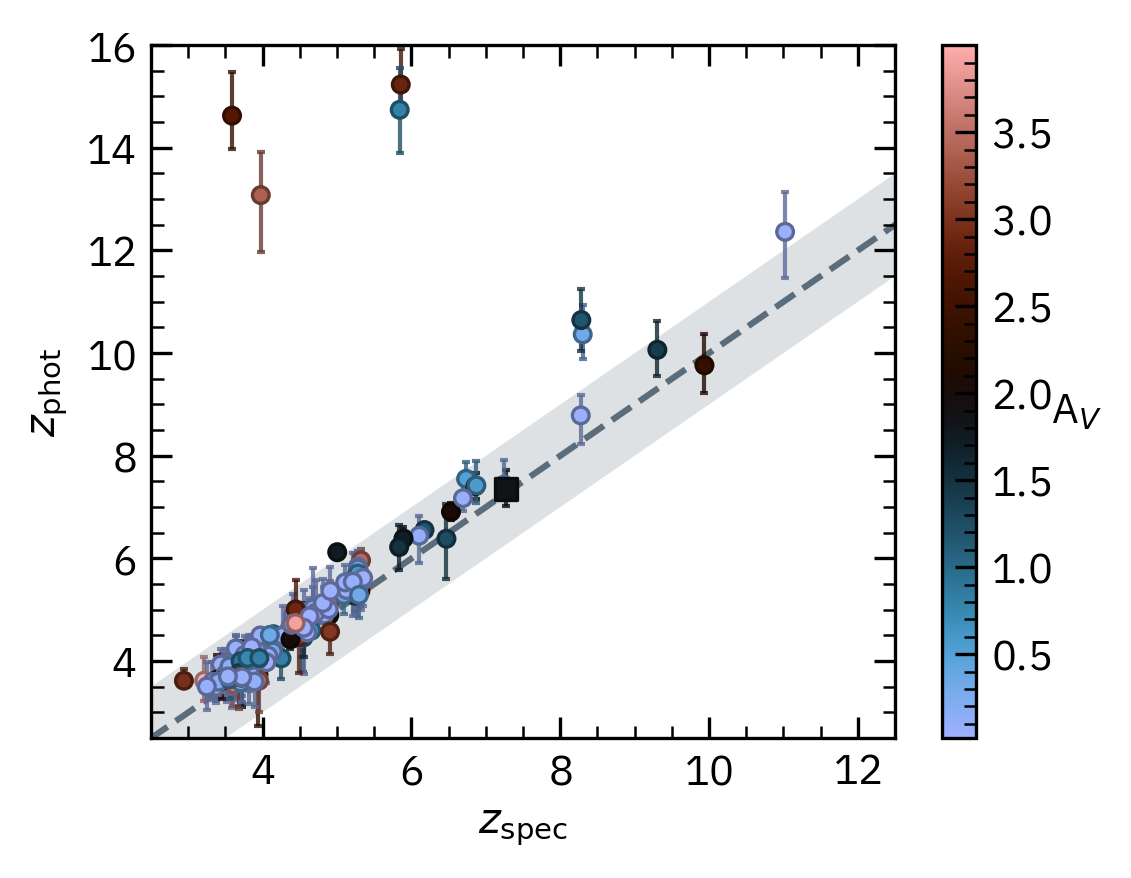}
    \caption{Comparison of photometric redshifts and confirmed spectroscopic redshifts for galaxies in the CAPERS sample, colored by $A_V$. Galaxies are selected to have photometric redshifts $z_{\mathrm{phot}} > 3.5$. The dashed line denotes the one-to-one relation with the shaded region having $\Delta z =$ 0.5. While the majority of sources show good agreement between photometric and spectroscopic redshifts, a small number of outliers are present, including four galaxies with $z_{\mathrm{phot}} > 12$ but $z_{\mathrm{spec}} < 7$, consistent with known degeneracies between high-redshift and dust-reddened lower-redshift photometric solutions, and were included in the target selection as likely $z\sim 4-6$ DSFGs.}
    \label{fig:z_comp}
\end{figure}

In Figure~\ref{fig:z_comp}, we show the comparison of photometric and confirmed spectroscopic redshifts (\S~\ref{sec:data}). Overall, we find strong agreement between photometric and spectroscopic redshifts across the full redshift range probed. Four galaxies (3\% of the full sample) have photometric redshifts \zphot$ > 12$ but confirmed spectroscopic redshifts \zspec$ < 7$. These sources were visually vetted during CAPERS target selection as likely massive dusty star-forming galaxies (DSFGs) at lower redshift $z \sim 4-7$ \citep{Zavala2023, ArrabalHaro2023Natur.622..707A} (with all sources having a $P(z)$ with a secondary peak at these redshifts), and were thus retained in the target sample. 
Aside from these cases, the photometric redshifts are broadly consistent with the spectroscopic redshifts, and the following spectro-photometric SED-fitting is performed with redshifts fixed to \zspec. 

\begin{table*}[t]
\centering
\setlength{\tabcolsep}{12pt}

\hspace*{-2.3cm}
\begin{tabular}{l c c c c}
\hline
Component & Parameter & Prior & Range & Additional Parameters \\
\hline\hline

\multicolumn{5}{c}{\citet{Iyer2019} Star-Formation History}  \\
\hline
Total stellar mass formed 
& $\log_{10}(M_\star/M_\odot)$ 
& Uniform & $(5,\ 14)$ \\

Star-formation rate 
& $\log_{10}$ SFR [$M_\odot\,\mathrm{yr}^{-1}$] 
& Uniform & $(-3,\ 4)$ \\

Stellar metallicity 
&  $\log_{10}$ $Z_\star/Z_\odot$ 
& Uniform & $(-3.0,\ 0.3)$ \\

Assembly timescales 
& $\tau_{25},\ \tau_{50},\ \tau_{75}$ 
& Free & $(0,\ t_\mathrm{univ})$ \\

\hline
\multicolumn{5}{c}{\citet{Salim2018} Dust Attenutation Law} \\
\hline
V-band attenuation 
& $A_V$ 
& Uniform & $(0,\ 4.5)$ \\

Deviation from Calzetti slope 
& $\delta$ 
& Uniform & $(-1.4,\ 0.6)$ \\

2175\,\AA\ bump strength 
& $B$ 
& Uniform & $(0,\ 4)$ \\

Nebular-to-stellar attenuation ratio 
& $\eta$ 
& Uniform & $(0,\ 2.5)$ \\

\hline
\multicolumn{5}{c}{Nebular Emission} \\
\hline
Ionization parameter 
& $\log_{10}(U)$ 
& Uniform & $(-4,\ 0)$ \\

Escape fraction of ionizing photons 
& $f_\mathrm{esc}$ 
& Logarithmic & $(10^{-3},\ 1)$ \\

\hline
\multicolumn{5}{c}{Ly$\alpha$ Damping-Wing Absorption} \\
\hline
Neutral hydrogen fraction 
& $x_\mathrm{HI}$ 
& Uniform & $(0,\ 1)$ \\

Neutral hydrogen column density 
& $\log_{10}(N_\mathrm{HI}/\mathrm{cm}^{-2})$ 
& Uniform & $(12,\ 24)$ \\

Effective damping-wing scale 
& $R$ 
& Uniform & $(0,\ 300)$ \\

\hline
\multicolumn{5}{c}{Noise Model} \\
\hline
White-noise scaling 
& $a$ 
& Logarithmic & $(0.1,\ 10)$ \\

\hline
\multicolumn{5}{c}{General Parameters} \\
\hline
Redshift 
& $z_\mathrm{obs}$ 
& Gaussian & $z_\mathrm{spec} \pm 0.25$ 
& $\mu=z_\mathrm{spec},\ \sigma=0.05$ \\

Stellar velocity dispersion 
& $\sigma_\star$ [$\mathrm{km\,s^{-1}}$] 
& Logarithmic & $(10,\ 1000)$ \\

\hline
\end{tabular}
\caption{Free parameters and associated priors used in our \bp\ spectro-photometric SED fitting. Logarithmic priors are base 10.  We describe the model in \S~\ref{subsec:sed_fitting}.}
\label{tab:BP_priors}
\end{table*}

\subsection{BAGPIPES Spectro-Photometric SED fitting} \label{subsec:sed_fitting}

To understand the formation timescales of these galaxies, we utilize \bp\ v1.3.3 \citep{Carnall2018, Carnall2019SFH}, jointly modeling photometry and spectra. We chose to utilize \bp\ as it is capable of jointly modeling  spectra and photometry efficiently. In \bp,
we use the Binary Population and Spectral Synthesis (BPASS) stellar models v2.2.1, using the 135\_300 IMF \citep{BPASS}.
The star-formation histories are parameterized according to the \db\ GPM, defining the same $\tau_{25}, \tau_{50},$ and $\tau_{75}$ shape parameters as discussed in \S \ref{subsec:sed_fitting_phot}. The stellar metallicity is allowed to vary with a uniform (in log-space) prior on the metallicity for log $Z/Z_\odot \in [-3, 0.3]$, and the total stellar mass of the galaxy at the epoch of observation is fitted with a uniform prior in logarithmic space over the range \logM $\in [5, 14]$, and log (SFR/$M_\odot$yr$^{-1}$) $\in [-3, 4]$, again with a logarithmic prior.

\begin{figure*}
\includegraphics[width=\linewidth]{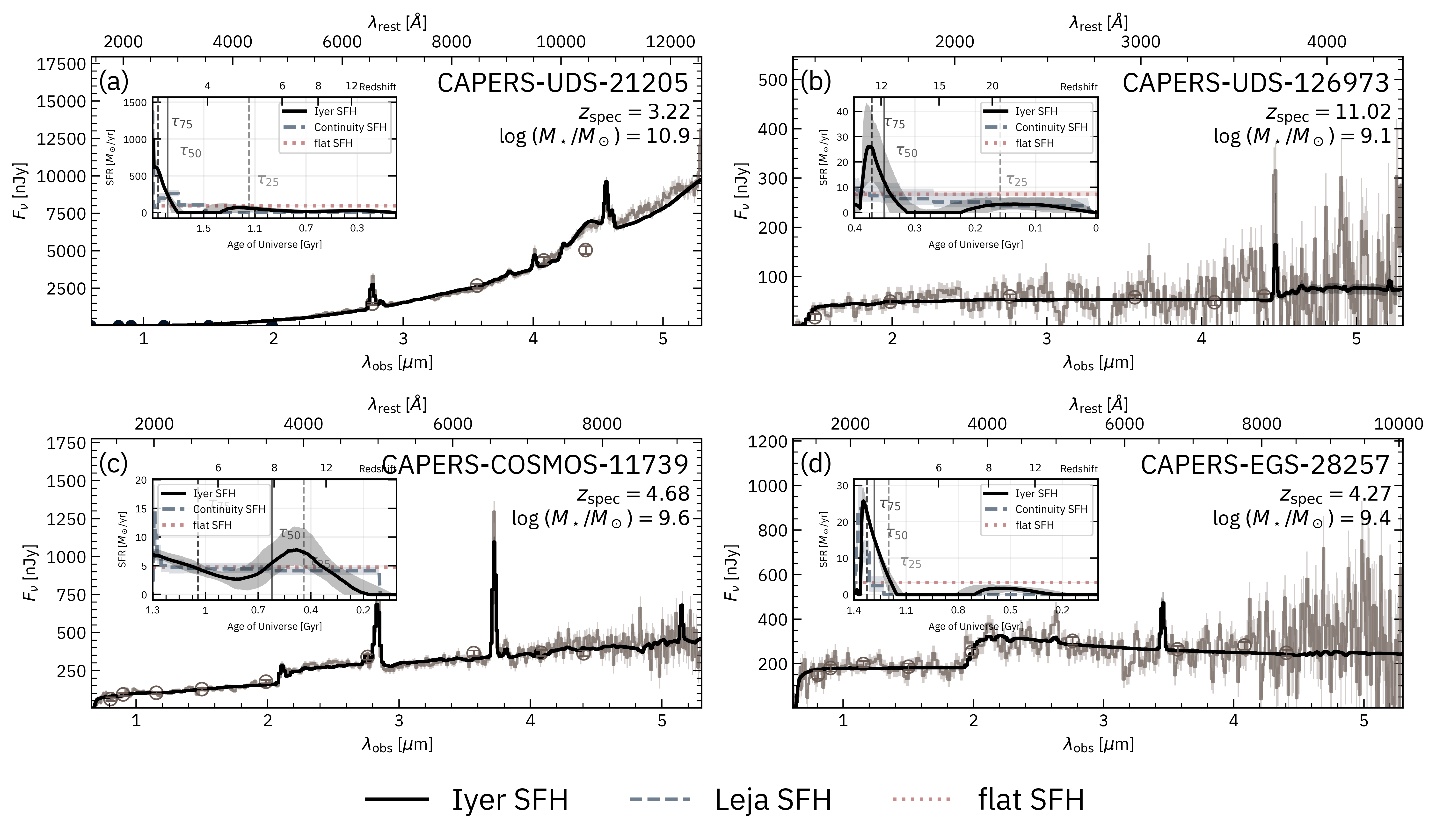}
\caption{Recovered SEDs from \bp\ fitting of joint spectroscopy and photometry of four galaxies in our sample. The data is shown in taupe (photometry shown as open circles with 1$\sigma$ errorbars) and the model fit in black. The shaded region shows the 1$\sigma$ error (spread) in the data (recovered fit). The inset in each figure shows the recovered SFHs using different SFH parameterizations: our fiducial\citet{Iyer2019} SFH in black, \citet{Leja2019} Continuity SFH in blue, and a constant SFH in red (see \S 4.6). The \citet{Iyer2019} SFH is the fiducial SFH used in the analysis of this work. Discussion of the SFHs can be found in \S~\ref{subsec:SFH_comp}. The vertical lines in the inset show time at which 25\%, 50\%, and 75\% of the stellar mass formed. Panel (a) shows the most massive galaxy in our sample,
Panel (b) is the highest redshift galaxy in our sample, and Panels (c) and (d) show the galaxy with the median stellar mass and redshift, respectively.  The complete figure set with all the galaxies analyzed (37 images) is available in the online journal.} \label{fig:SED_SFH}
\end{figure*}

We allow for flexible dust attenuation using the model of \citet{Salim2018} (see also \citealt{Noll2009}), which parametrizes the shape of the dust curve in terms of a power-law deviation, $\delta$, from the \citet{Calzetti2000} local starburst dust law (defined as having $\delta = 0$). We set a uniform prior on $\delta \in (-1.4, 0.6)$, we found that this broad prior is necessary to provide adequate fits to the spectra examined here. Additionally, we set a uniform prior on the strength of the 2175\AA\ bump, $B,$ from $(0, 4)$, where the Milky Way law has $B$ = 3. We allow $A_V$ to vary between 0 and 5 uniformly, and the ratio of stellar to nebular dust attenuation ($\eta$) to vary between 0 and 2.5.
The nebular line and continuum emission is determined by post-processing the BPASS stellar models using the Cloudy photoionization code \citep{BPASS,CLOUDY}, with the nebular metallicity equal to the stellar, and with an ionization parameter $U$ allowed to vary between log $U \in [-4, 0]$ in steps of 0.5. The escape fraction of ionizing photons ($f_\mathrm{esc}$) is allowed to vary with a uniform logarithmic prior between (0.001,1). We additionally include a Ly$\alpha$ damping-wing absorption component to capture attenuation from neutral hydrogen along the line of sight. This component is fit with uniform priors on the neutral fraction, $x_{\mathrm{HI}} \in [0,1]$, an effective scale parameter $R \in [0,300]$, and the neutral hydrogen column density $\log N_{\mathrm{HI}} \in [12,24]$ (in cm$^{-2}$). Example SED fits are shown for four objects in Figure~\ref{fig:SED_SFH}; the fits for the complete galaxy sample is available in the online journal. The physical properties of our final sample of 148 galaxies as derived from the spectro-photometric SED  are presented in Table~\ref{tab:capers_prop}.


\figsetstart
\figsetnum{3}
\figsettitle{SED fits and SFHs of all galaxies}

\figsetgrpstart
\figsetgrpnum{3.1}
\figsetgrptitle{Galaxy fits}
\figsetplot{SED+SFH_panel_1.png}
\figsetgrpnote{Recovered SEDs from \bp\ fitting of joint spectroscopy and photometry of four galaxies in our sample, ordered by stellar mass. The data is shown in taupe (photometry shown as open circles) and the model fit in black. The inset in each figure shows the recovered SFHs using different SFH parameterizations. The Iyer SFH, shown in black, is the fiducial SFH used in the analysis of this work. Discussion of additional SFHs can be found in \S~\ref{subsec:SFH_comp}. 
The vertical lines in the inset show time at which 25\%, 50\%, and 75\% of the stellar mass formed as determined from our fiducial model.}
\figsetgrpend

\figsetend

\setcounter{figure}{3} 

\section{Properties of Galaxies} \label{sec:analysis}

\subsection{Stellar Mass Recovery} \label{subsec:Mstar}

\begin{figure}[th]
    \centering
    \includegraphics[width=\columnwidth]{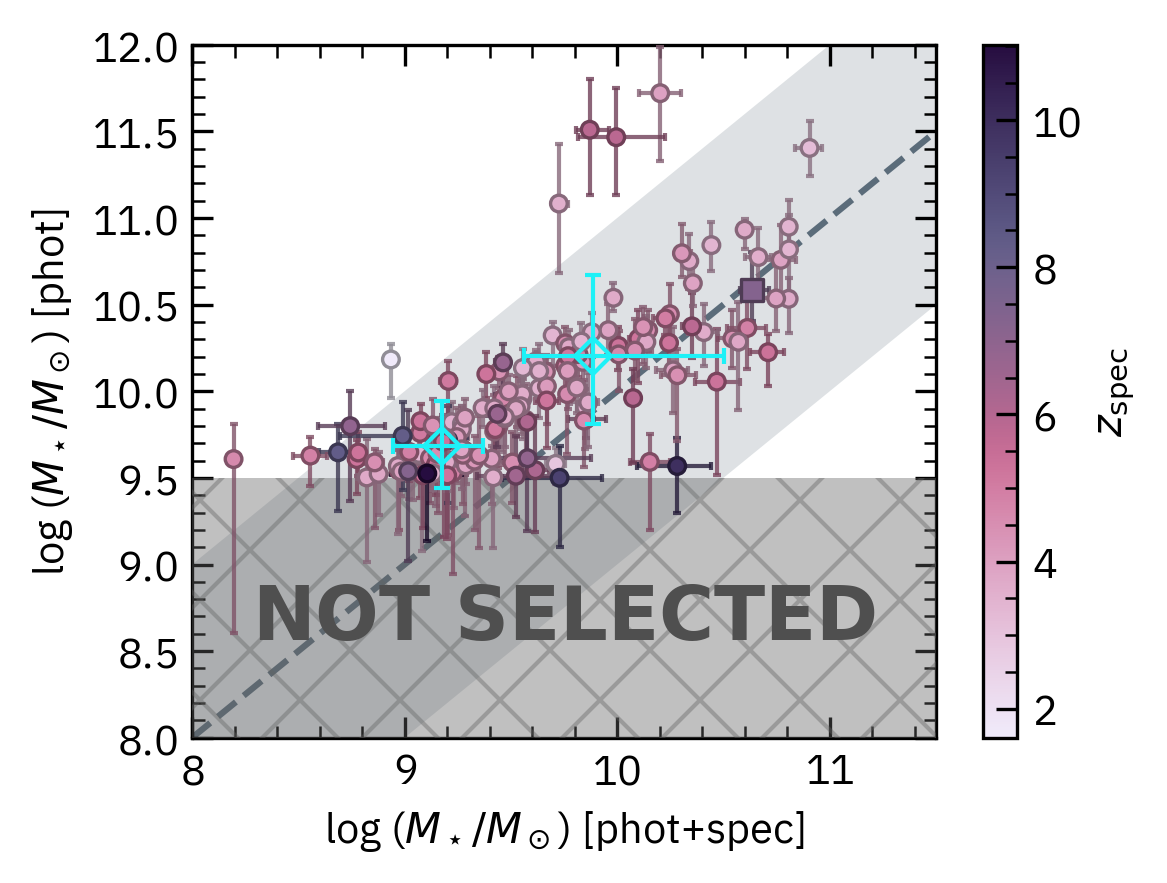}
    \caption{Comparison of stellar mass estimates derived from joint spectroscopic and photometric SED fitting and from photometry-only SED fitting, colored by redshift. Galaxies in our sample were selected to have photometric stellar masses of $\log M_\star > 9.5$. The open cyan points indicate the median stellar masses in two bins defined by the stellar mass inferred from joint spectro-photometric SED fitting: $\log M_\star < 9.5$ and $\log M_\star \geq 9.5$. The dashed line is a 1-to-1 relation while the shaded region shows a $\pm$1 dex range. While overall agreement is good, a small deviation towards higher photometric masses is likely due to the misrepresentation of nebular emission as stellar continuum emission in broad band filters, which can be better decoupled with the inclusion of spectroscopy. Additionally, owing to the photometric mass selection, the sample is incomplete below \logM $= 9.5$, which biases the median in the lower-mass bin and causes it to deviate from the one-to-one relation.}
    \label{fig:Mstar_comp}
\end{figure}

Figure \ref{fig:Mstar_comp} compares the stellar masses inferred from photometry alone from \db\ and those inferred from joint photometric and spectroscopic SED fitting with \bp. We calculate the median inferred stellar masses in two bins, one for spectro-photometric \logM $< 9.5$ and another at \logM $\geq 9.5$. 
For both mass bins, photometry-only fits tend to yield systematically higher stellar mass estimates of 0.51 and 0.29 dex in the lower and higher mass bin, respectively. 
Additionally, our sample is largely incomplete (based on our photometric mass selection) below $\log M_\star = 9.5$, further biasing high the median photometric inferred stellar mass of the lower mass bin.

\begin{figure*}[ht]
    \centering
    \includegraphics[width=\textwidth]{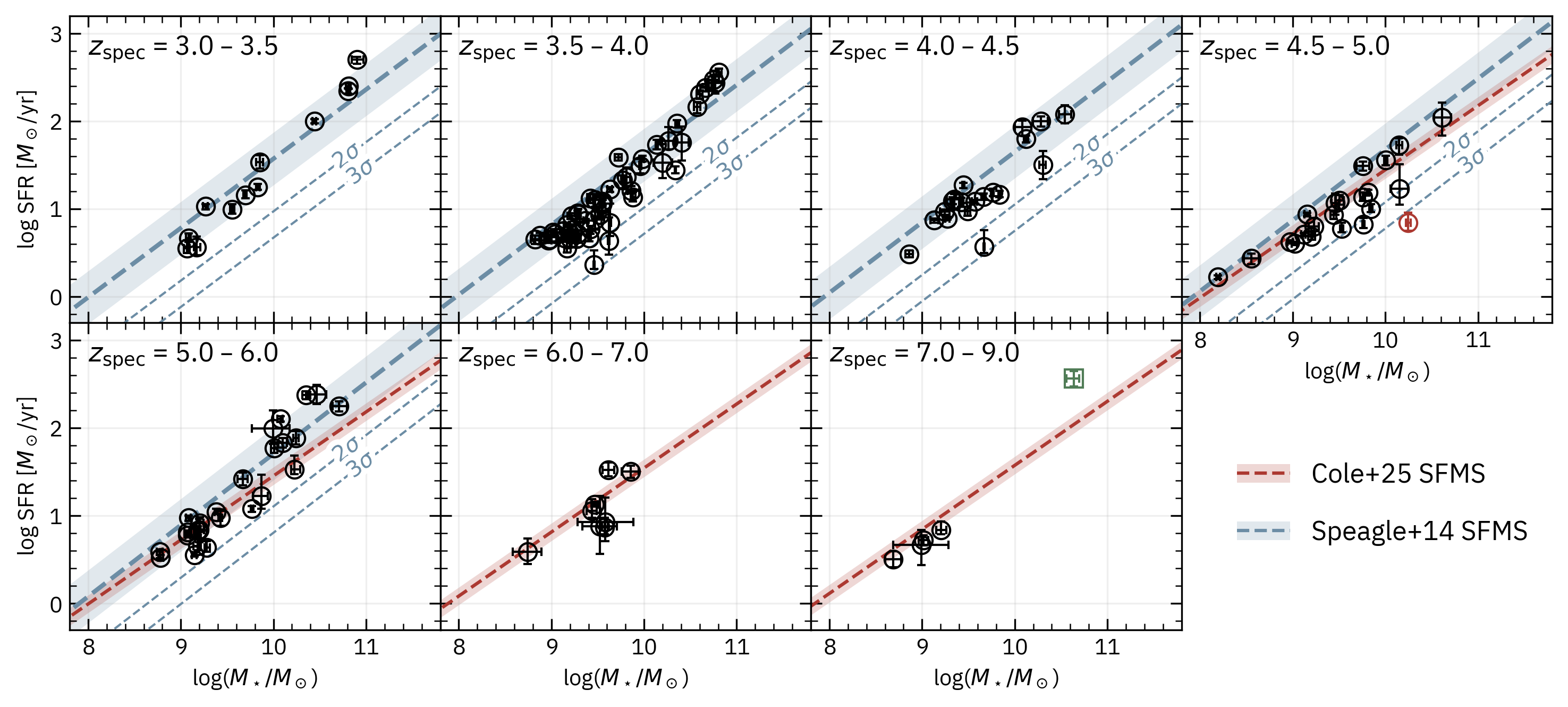}
    \caption{Stellar mass versus star-formation rate (averaged over 100~Myr) for our galaxy sample, shown in redshift bins spanning $3 < z < 9$. In each panel, galaxies are compared to published star-forming main-sequence (SFMS) relations. The \citet{Speagle2014} SFMS is shown in blue for panels at $z=3$--6, while the \citet{Cole2025} SFMS is shown in red for panels at $z=4.5$--9. The shaded region shows the 1$\sigma$ spread in these relations. The majority of galaxies lie on or near the SFMS at their respective redshifts, consistent with typical star-forming systems at these epochs. We find one quiescent galaxy (red circle), classified as having a SFR more than $3 \sigma$ below the SFMS at their masses and redshift. The \citet{Weibel2025} massive quiescent galaxy is shown as a green square. We did not classify this galaxy as quiescent as the SFH recovery showed a significant burst of star-formation $\sim 20$ Myr before the time of observation, followed by rapid quenching, which is not captured by our 100 Myr averaged SFR measurement.}
    \label{fig:SFMS}
\end{figure*}

Similar trends have been reported in previous studies comparing photometric and spectro-photometric SED fitting \citep[e.g.][]{PerezGonzalez2023, Hamed2026}. 
The overall offset between photometric and spectro-photometric stellar mass estimates is likely due to the contribution of strong nebular emission lines, which are often not accurately constrained by broadband photometry alone. In star-forming galaxies, prominent rest-frame optical emission lines can contribute substantial flux to broadband filters, boosting the inferred continuum flux when spectra are not available \citep[e.g.][]{Salmon2015, Pacifici2023, Cochrane2025, Chary2005, Shim2011}. In the absence of spectroscopic constraints, this excess flux can be misattributed to stellar continuum emission, leading to systematically overestimated stellar masses. This effect is expected to be most pronounced at lower stellar masses, where the underlying stellar continuum is weaker and nebular emission can dominate the broadband flux. Our results are therefore consistent with previous findings that photometry-only stellar masses in low-mass, strongly star-forming galaxies may be biased high when nebular emission is not explicitly accounted for. Including spectroscopic data enables SED fitting to explicitly separate continuum and line emission, resulting in more accurate stellar mass estimates. 
We note, however, that our strict selection cut at photometric stellar masses of \logM$>9.5$ limits our ability to robustly quantify the magnitude of this offset at lower masses since we do not consider galaxies with lower stellar masses.

\subsection{$M_\star$ vs. Star-Formation Rate} \label{subsec:SFMS}
Figure~\ref{fig:SFMS} shows the stellar mass versus the star-formation rate (where the latter is the value time-averaged over the past 100 Myr from the observed epoch) for our galaxy sample, compared to published star-forming main sequences \citep[SFMS;][]{Cole2025, Speagle2014} for seven redshift bins spanning $ 3 < z < 9$. The vast majority of our galaxies lie on or near the SFMS at their corresponding redshifts, consistent with typical star-forming galaxies at these epochs.  

We find a single galaxy considered massive and quiescent, defined as having SFR $>3 \sigma$ below the \citet{Speagle2014} SFMS at its stellar mass and redshift: CAPERS-EGS-21360, having a stellar masses of \logM=$10.31 \pm 0.02$, and sSFR of $-9.43 \pm -0.09  \ \textrm{yr}^{-1}$ at \zspec$=4.55$. RUBIES-UDS-QG-z7 is in our analyzed sample, this galaxy was first identified as a massive quiescent galaxy by \citet{Weibel2025}. Our SFH recovery showed significant star-formation $\sim 20$ Myr before the time of observation, followed by rapid quenching. This rapid quenching was not recovered by 100 Myr averaged SFR measurement and thus not classified as quiescent in this work. Additional discussion of quiescent galaxies in the entire CAPERS survey can be found in Shen et al. (submitted). 


\subsection{Deviation from Calzetti Dust Attenuation} \label{subsec:delta}

\begin{figure}[ht]
    \centering
    \includegraphics[width=0.95\linewidth]{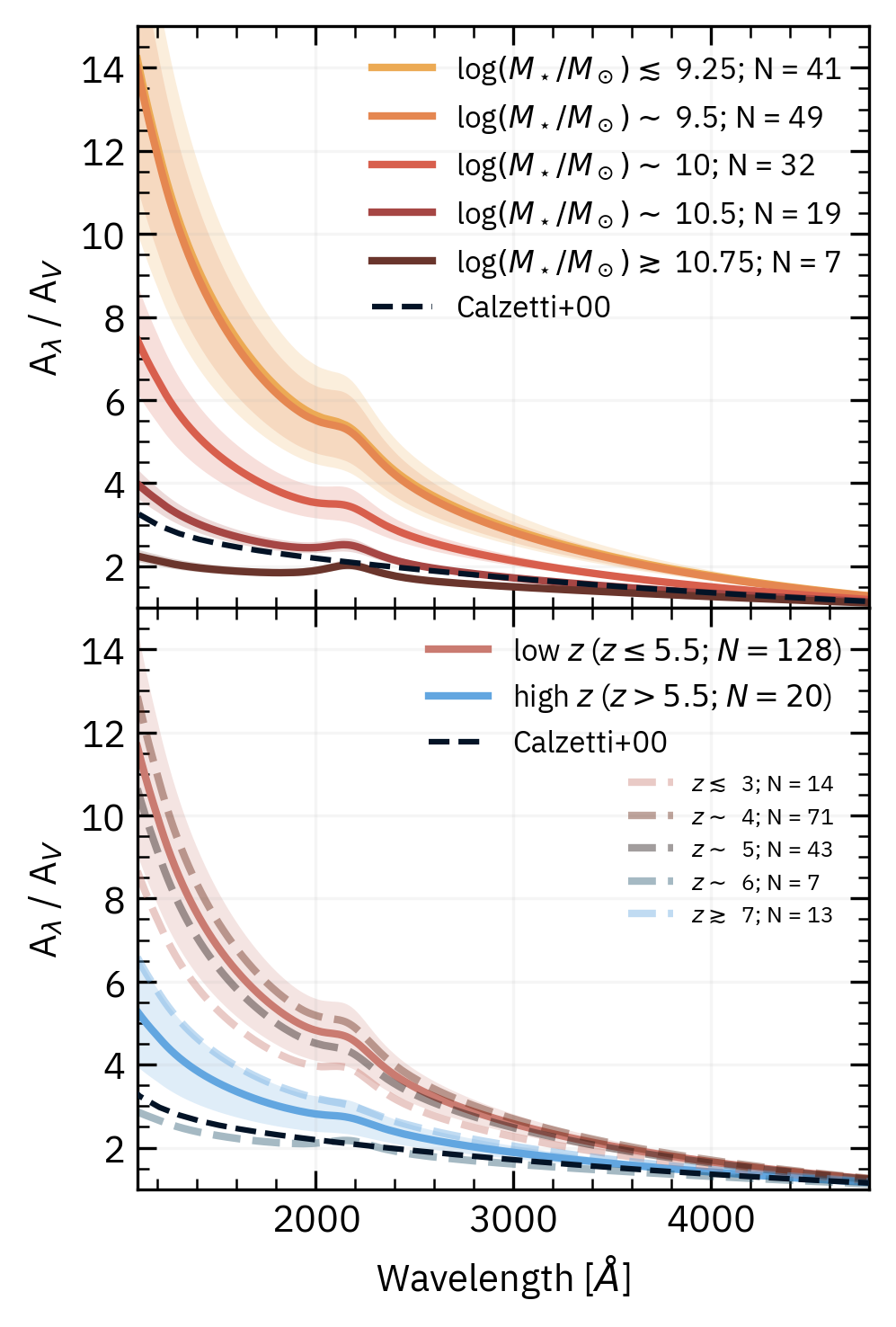}
    \caption{Recovered attenuation curves of our sample of massive galaxies, binned by stellar mass (Top) and redshift (bottom). \textit{Top: } The lines show median attenuation curves and the shaded region in the 1$\sigma$ spread. We see that more massive galaxies tend to exhibit shallower attenuation curves, while lower mass galaxies tend to have stepper attenuation curves with larger scatter. \textit{Bottom}: We separate our full galaxy sample into a low- ($z<5.5$) and high- ($z \geq 5.5$) redshift sample.  The dashed lines show the median attenuation curve in finer redshift bins. The bulk of our sample of massive galaxies fall at a redshift of $z<5.5$ (Figure~\ref{fig:sample_dist}), and there exists minimal evolution in slope for these galaxies, however, at $z>5.5$ we see an evolution towards shallower attenuation curves. }
    \label{fig:Av_binned}
\end{figure}

\begin{figure*}[t]
    \centering
    \includegraphics[width=\textwidth]{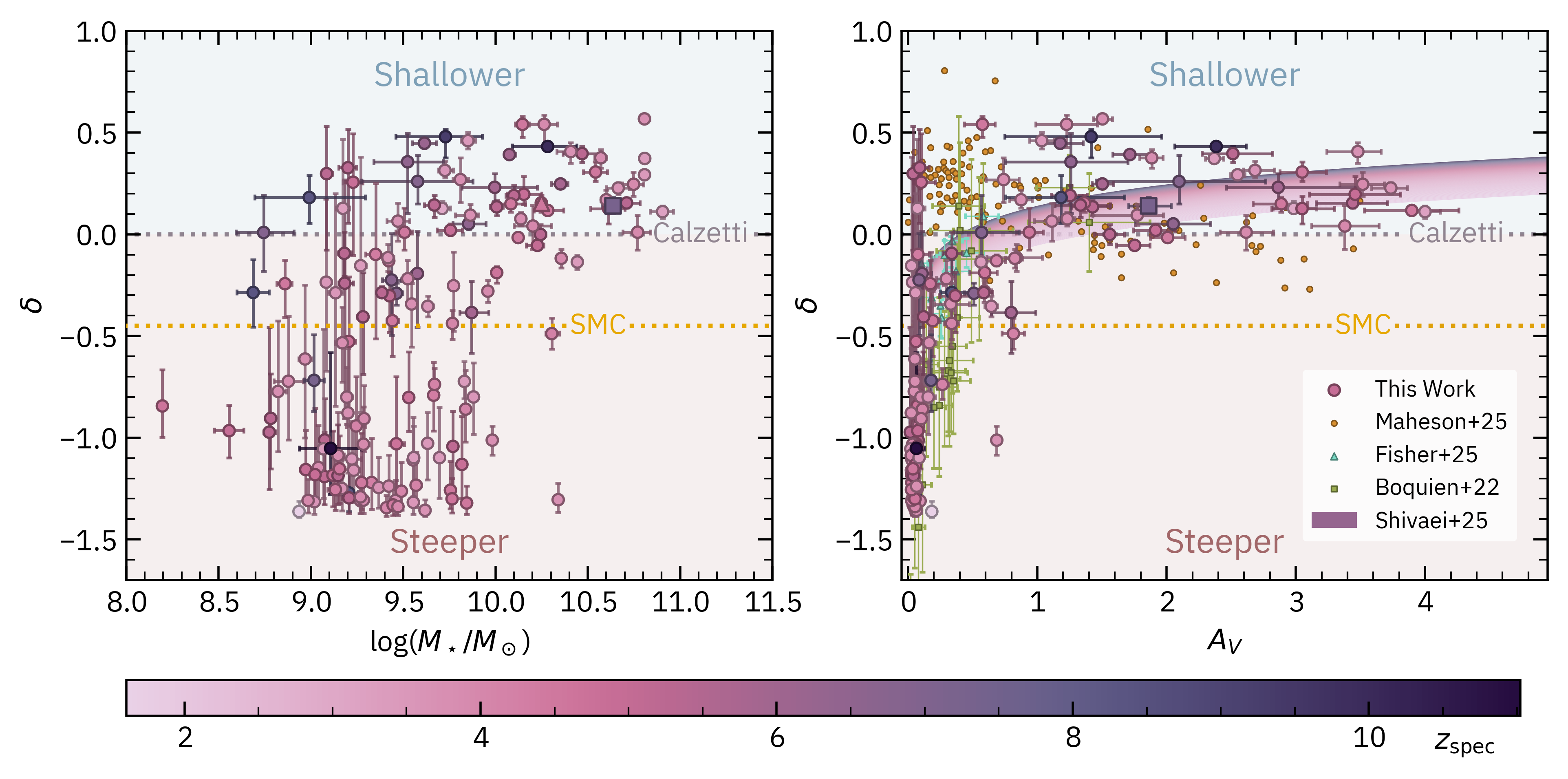}
    \caption{\textit{Left:} Distribution of the dust attenuation slope parameter $\delta$ as a function of stellar mass. More massive galaxies preferentially exhibit shallower than Calzetti attenuation curves ($\delta > 0$), with nearly all galaxies above $M_\star \sim 10^{10} M_\odot$ having attenuation curves flatter than the SMC law. Lower-mass galaxies span a broader range of $\delta$, including steeper, SMC-like attenuation.
    \textit{Right:} Dust attenuation slope $\delta$ as a function of total attenuation $A_V$. The orange circles are results from lower redshift ($1.7 < z< 3.5$) massive galaxies ($9 <$ \logM $<11.5$) from the \textit{JWST} Blue Jay survey \citep{Maheson2025}. The cyan triangles are $z\sim7$ massive  ($9 <$ \logM $<10$) galaxies from the REBELS-IFU survey \citep{Fisher2026}, and the green squares show dust properties as determined from the ALPINE-ALMA [C \textsc{ii}] survey at $z \sim 5$ \citep{Boquien2022}. The shaded pink curve shows evolution of $\delta$ from $z=1-9$ as determined from \textit{JWST} NIRCam observations of $\sim 3,800$ galaxies \citep{Shivaei2025}. There is a weak dependence on redshift, with higher redshift galaxies having shallower attentuation curves compared to the bulk of the population.
    } 
    \label{fig:delta}
\end{figure*}

These galaxies were fit with the \citet{Salim2018} dust attenuation law, a modification of the \citet{Calzetti2000} law, allowing for a more flexible attenuation curve. In the \citet{Salim2018} dust law, the slope of the Calzetti curve is parametrized by a power-law term having an exponent $\delta$, where the Calzetti curve has $\delta = 0$. A negative (positive) $\delta$ produces a steeper (shallower) attenuation curve. A steeper curve implies greater relative attenuation at shorter wavelengths, consistent with simpler foreground dust-screen geometries. Whereas a grayer curve suggests more complex, mixed dust-star geometries \citep{Salim2020, Burgarella2025A&A...699A.336B}.
An additional parametrization, $B$, adds a 2175\AA\ bump following a Drude profile \citep{FitzpatrickMassa1986}. We also incorporate an $\eta$ term, allowing for differential attenuation between stellar continua and nebular emission. While fitting our sample of galaxies, we found that a large range of $\delta$ was required to simultaneously recover the range of slopes in the rest-UV continuum as well as the strength of the emission lines (in particular, the ratios between the Balmer lines).
We show the recovered attenuation curves of our galaxies binned by stellar mass and redshift in Figure~\ref{fig:Av_binned}, and in Figure~\ref{fig:delta} we show the spread of $\delta$ compared to the galaxies' stellar mass and $A_V$. We find that the more massive galaxies in our sample tend to have shallower dust attenuation curves ($\delta > 0$), with nearly all galaxies with stellar mass above $10^{10.5} M_\odot$ having $\delta > 0$. In comparison, at stellar masses below $10^{10} M_\odot$, we find substantial scatter in the recovered $\delta$ values.  We find that the scatter in $\delta$ is largest at $A_V \lesssim 1$. This may be due, in part, that at low reddening it becomes increasingly difficult to constrain the shape of the attenuation curve.. 

Since stellar mass and dust attenuation are physically linked, as more massive galaxies have undergone more star formation and thus produced more dust \citep{GarnBest2010, Pannella2009, Salim2020}, the observed trends in $\delta$ with both mass and $A_V$ likely reflect the same underlying increase in dust column density and geometric complexity toward more massive systems. 
Galaxies at higher redshifts also appear to have shallower dust attenuation curves, where galaxies with spectroscopic redshifts $z>6$ all have attenuation curves shallower than the Small Magellenic Cloud (SMC; $\delta = -0.45$; \citet{Gordon2003}), and a majority of these galaxies having attenuation curves shallower than the Calzetti law, consistent with results presented by \citet{Markov2025}. 



\subsection{Formation Timescales of Galaxies} \label{subsec:tau}

\begin{figure*}[ht]
    \centering
    \includegraphics[width=\textwidth]{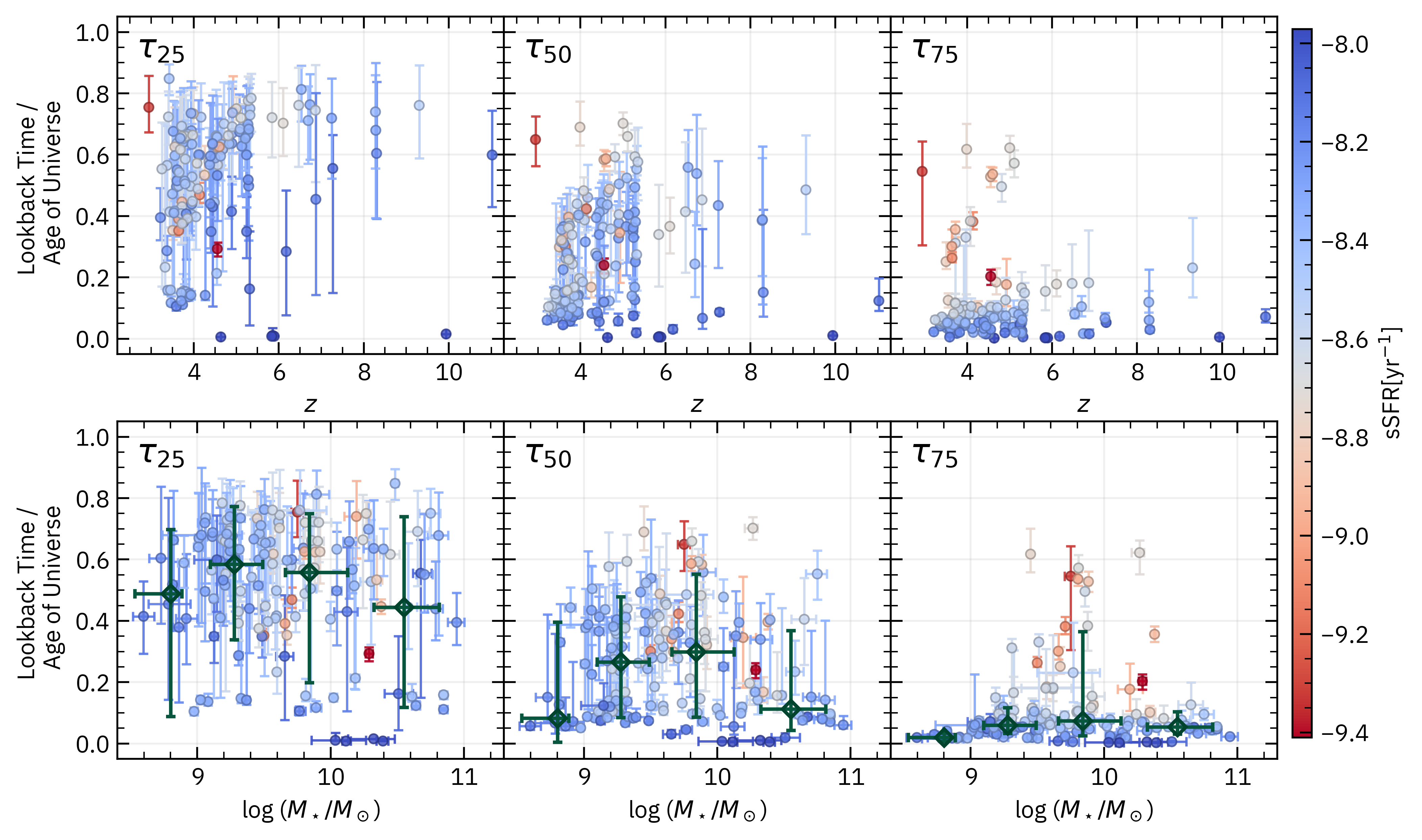}
    \caption{SFH assembly timescales plotted against redshift (top) and stellar mass (bottom); the points are colored by sSFR. Shown are the fractional lookback times at which galaxies reach key assembly epochs (where 0 represents the epoch of observation and 1 the Big Bang), normalized to the age of the Universe at the epoch of observation.
    The top panel shows how the SFH assembly timescales evolve with redshift; we find that lower-redshift galaxies show large scatter in the onset of significant star formation ($\tau_{25}$). The higher-redshift sources tend to form early in the Universe's lifetime, at later formation epochs ($\tau_{50}$, $\tau_{75}$) assembly timescales converge across redshifts.  However, we note that conclusions are  limited by the small number of galaxies at $z \gtrsim 6$ in this work. 
    The bottom panel shows SFH assembly timescales with respect to stellar mass. The diamonds show the median relations in bins of stellar mass with the error bars denoting the 16th–84th percentile spread.
    From both panels, we see that galaxies with low sSFR preferentially populate the high $(t_\mathrm{Univ} - \tau_x)/t_\mathrm{Univ}$ tail, indicating early completion of stellar mass growth and suppressed star formation at the epoch of observation.}
    \label{fig:tau_obs}
\end{figure*}

\begin{figure*}[ht]
    \centering
    \includegraphics[width=\textwidth]{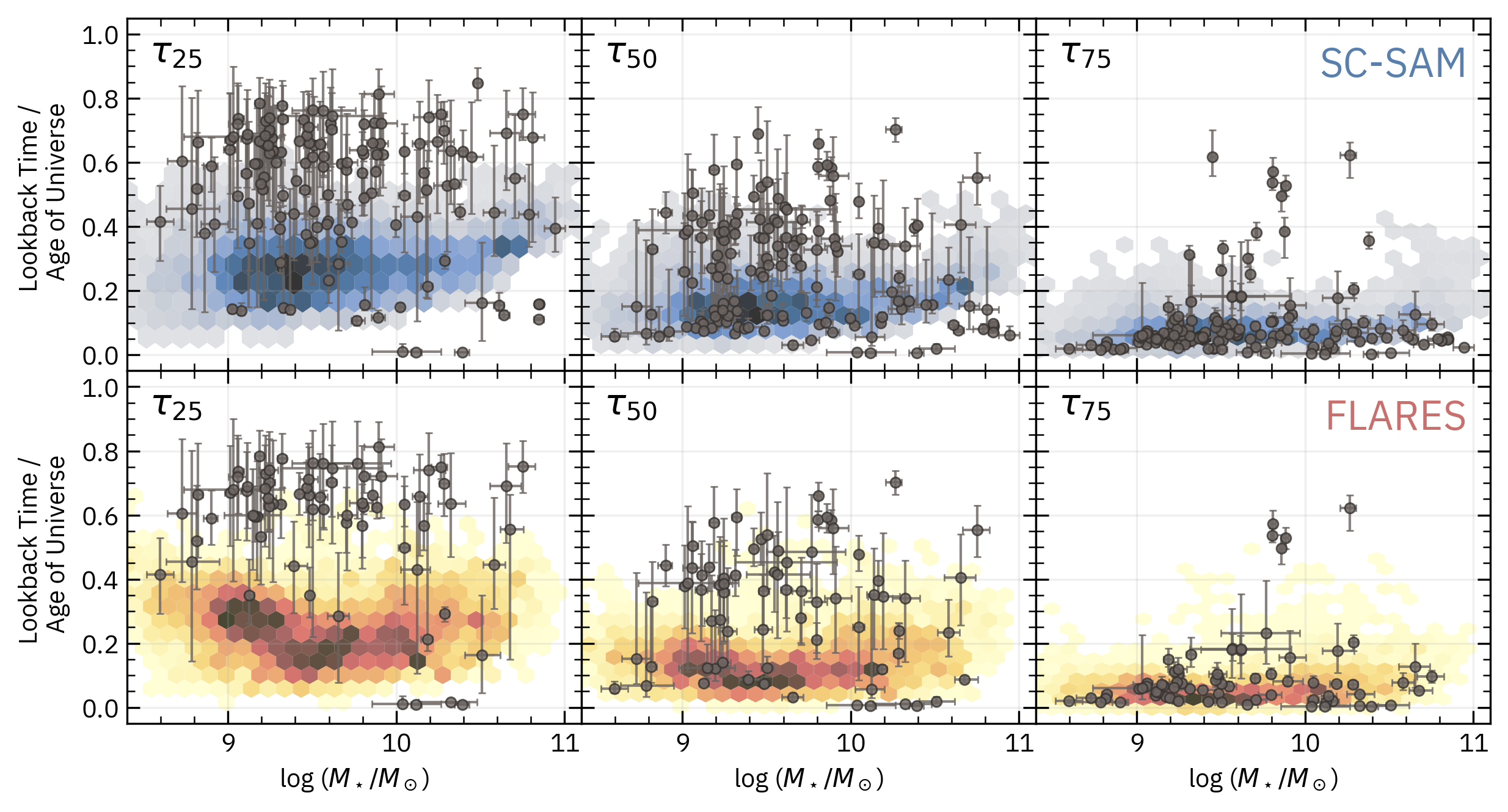}
    \caption{
    The top panel compares our observations (gray circles) to galaxies from an $\sim 800$ arcmin$^2$ lightcone with SC-SAM galaxy formation physics \citep[shaded hexagons;][]{Yung2022}. The bottom panel shows comparisons to the FLARES simulations \citep{Lovell2021}, a suite of high-resolution, hydrodynamic simulations using the EAGLE \citep{EAGLE2017} physics. For each panel, we only show observed galaxies that have simulation counterparts with similar mass and redshifts (details in \S~\ref{subsec:tau}). As the FLARES simulation is run to $z=4.77$, there are fewer matches to our observed sample of galaxies. The early assembly period ($\tau_{25}$) of observed galaxies is not well reproduced by either the SC-SAM or FLARES galaxies, however, agreement improves towards later galaxy epochs, suggesting that the onset of star-formation in these simulations occur at a later time than what is observed.}
    \label{fig:tau_sims}
\end{figure*}

To understand the formation timescales of the galaxies in this work, we explore the dependences of $\tau_x$, or time at which galaxies formed $x$\% of their stellar mass, with respect to stellar mass and redshift.
Figure~\ref{fig:tau_obs} shows the fractional lookback times at which galaxies reached 25\%, 50\%, and 75\% of their stellar mass at the epoch of observation, expressed as $(t_\mathrm{univ} - \tau_x)/t_\mathrm{univ}$, where $t_\mathrm{univ}$ is the age of the Universe at the epoch of observation. This normalization removes the redshift dependence associated with the younger age of the Universe at high redshift and enables a comparison of relative assembly histories across the sample.  
For the early assembly timescale, $\tau_{25}$, while the median is relatively constant across stellar mass, the large scatter in $\tau_{25}$ for this sample of galaxies indicates that the time at which the onset of significant star formation occurs differs significantly. 
The median epoch at which galaxies form $75 \%$ of their stellar mass is uniform across the entire mass range probed here. We find no clear redshift dependence in the normalized formation metrics aside from increased scatter at lower redshifts (Figure~\ref{fig:tau_obs}, \textit{top}), indicating that the diversity in inferred SFH shapes is not primarily driven by cosmic epoch across the redshift range probed. However, we find that galaxies with low sSFR systematically occupy the high end of the $(t_\mathrm{Univ} - \tau_x)/t_\mathrm{Univ}$ distribution, for $\tau_{50}$ and $\tau_{75}$, typically lying beyond $1\sigma$ of the population median.  
As expected, galaxies which have lower sSFRs, or have shut down their star-formation, assemble the majority of their stellar mass earlier than the typical population. 

In Figure~\ref{fig:tau_sims} we show comparisons of observed galaxy formation epochs to that determined for simulated galaxies in the mock $\sim 800$ arcmin$^{2}$ lightcone presented in \citet{Yung2022,Yung2019, Yung2019MNRAS.490.2855Y}, which utilizes the Santa Cruz semi-analytic model (SC-SAM) galaxy formation physics \citep{SomervillePrimack1999, Somerville2008, Somerville2012,Somerville2015,Somerville2021} (\textit{top}), and the FLARES suit of 40 `zoom' hydrodynamic simulations, based on EAGLE physics \citep{Lovell2021, EAGLE2017} (\textit{bottom}). To obtain a population of simulated galaxies, for each observed galaxy with stellar mass $M_\star$ and $z_\mathrm{spec}$, we pull 100 galaxies from each simulation with $\Delta M_\star < 0.25$ and $\Delta z < 0.05$ for the SC-SAM lightcone and $\Delta z < 0.5$ for FLARES (as the SC-SAM lightcones are more finely sampled over redshift space). We then calculate the formation timescales $\tau_{25}, \tau_{50}$, and $\tau_{75}$ from the simulated galaxy SFHs. We see that the epoch of significant star-formation ($\tau_{25}$) largely differs from that inferred from observations, with both the SC-SAM and FLARES galaxies forming at a later time than observed galaxies across the entire range of stellar masses. At the later formation epochs, $\tau_{50}$ and $\tau_{75}$, the predictions begin to converge with observations, with $\tau_{75}$ from both simulations being largely consistent with $\tau_{75}$ of the observed galaxies.

We also examine the star-formation histories of each galaxy individually. While a range of star-formation history shapes is recovered, the majority of the sample exhibits a recent burst of star formation within the past 10 Myr. We define a galaxy as having a recent burst if either (1) it is currently experiencing the largest burst of star formation (i.e., the maximum value across its entire star-formation history), or (2) the average star-formation rate over a 10 Myr window is at least $3\times$ the average star-formation rate over the galaxy’s entire lifetime. Using this criterion, we find that 113 of the 148 galaxies in our sample satisfy the definition of a recent burst. The prevalence of recent bursts indicates that the observed sample is dominated by galaxies undergoing significant star formation at the epoch of observation.

Together, these results suggest that current theoretical models may underestimate the efficiency or earliness of initial star-formation activity in these galaxies, while the cumulative later growth converges with predictions. The high fraction of recent bursts further implies that bursty star-formation modes, may play a critical role in shaping the observed properties of this population \citep{Endsley2023, Looser2024}.


\subsection{Effects of Slitloss Rescaling} \label{subsec:slitloss}

For the SED-fitting performed for this work, we perform a slitloss rescaling to manually match spectral flux levels to the photometric catalog (\S~\ref{subsec:spectra_norm}). This step compensates for wavelength-dependent throughput losses and aperture mismatches between the spectroscopic extraction and the total photometric flux. To assess the impact of this correction on derived physical parameters, we perform SED fits on the spectra both before and after slitloss rescaling with the same priors. 

\begin{figure*}[th]
    \centering
    \includegraphics[width=\textwidth]{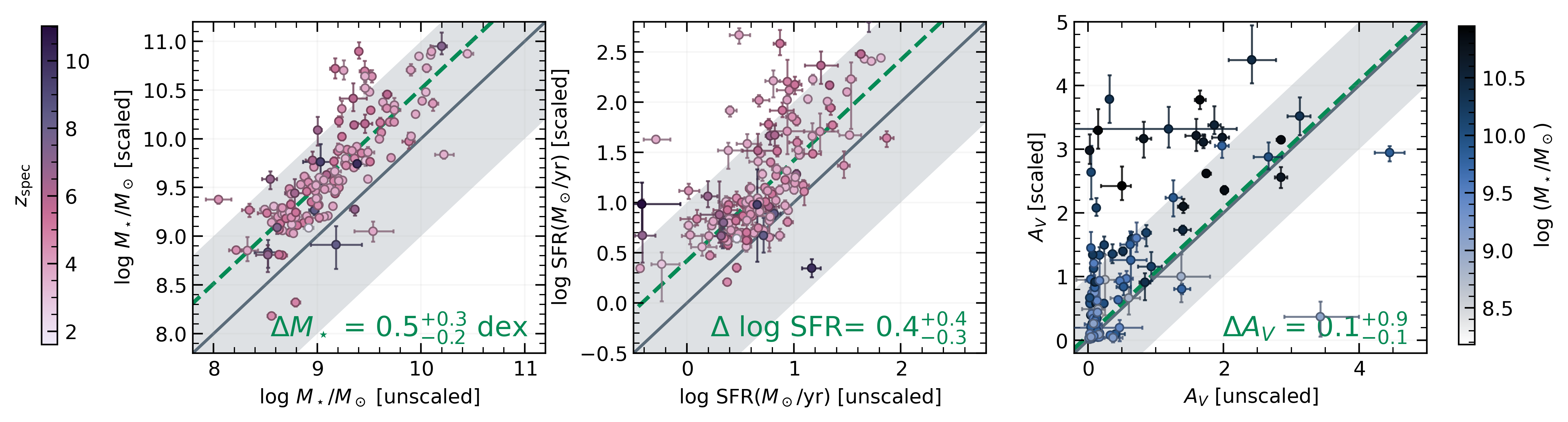}
    \caption{The effect of slitloss rescaling on the physical properties inferred from spectro-photometric SED fitting. Shown are stellar mass ($M_\star$; left), star-formation rate (SFR; middle), and dust attenuation ($A_V$; right), where the x-axis corresponds to values inferred from unscaled spectra and the y-axis to values inferred after applying a wavelength-dependent slitloss rescaling that matches the integrated spectroscopic flux to the broadband photometry (\S~\ref{subsec:spectra_norm}). Individual galaxies are shown as points colored by spectroscopic redshift, with error bars indicating the 16th–84th percentile spread from the SED fitting. The solid lines denote the one-to-one relation, and the shaded regions indicate deviations of $\pm1$. The dashed lines show the median offset between the two fits. Stellar masses and SFRs inferred from slitloss-rescaled spectra are systematically higher by $\sim0.5$ dex and $\sim0.4$, respectively, reflecting the increase in continuum and line flux normalization required to match the photometry. In contrast, the inferred dust attenuation exhibits larger object-to-object variation, with a small subset of sources showing substantial changes, consistent with the sensitivity of $A_V$ to wavelength-dependent differences between the spectral shape and broadband photometry. The effects of slitloss rescaling on physical properties is largely as expected, and we argue that for the properties of interest in this paper, performing this correction is best practice.
}
    \label{fig:scaled_props}
\end{figure*} 

Figure~\ref{fig:scaled_props} compares the inferred stellar mass, SFR, and dust attenuation ($A_V$) for the slitloss-rescaled and unscaled fits. We find that both $M_\star$ and SFR inferred from rescaled spectra are systematically higher by $\sim 0.5$ and $\sim 0.4$ dex, respectively. This offset is expected as slitloss rescaling typically scale the continuum and emission-line flux upward to match the photometric flux level, and the increased continuum normalization propagates directly into higher inferred stellar masses, while larger nebular line fluxes (and/or larger UV continuum normalization) drive higher inferred SFRs. The majority of sources remain within $\pm 1$ dex of the one-to-one relation for both $M_\star$ and SFR, indicating that slitloss primarily introduces a near-uniform rescaling rather than qualitatively changing the inferred properties for most galaxies.

The dust attenuation shows substantially larger variation. While the median $\Delta A_V$ is only $\sim 0.1$, a small subset of galaxies show large differences between the two fits, with $\Delta A_V > 1$ for 18 sources and a few sources having $\Delta A_V \sim 3$. 
This effect is likely a result of $A_V$ being constrained primarily by the spectral energy distribution shape, rather than its absolute normalization. In our calibration procedure, the slitloss rescaling is implemented as a wavelength-dependent multiplicative factor parametrized by a Chebyshev polynomial. While this approach preserves the integrated flux scale, it can introduce modest wavelength-dependent reshaping of the continuum and alter the relative weighting of emission-line–dominated and continuum-dominated spectral regions. Such effects are partially degenerate with dust attenuation and can therefore lead to significant changes in the inferred $A_V$, particularly for sources with strong emission lines or non-uniform dust and line-emitting geometries.

We argue that for the properties of interest in this paper, scaling the spectral flux to match the underlying photometry is the best practice. Stellar masses would not be accurately inferred without taking into account the entire light contribution of the galaxy. Similarly, the overall color of the galaxy has major contributions to the inferred dust content. For the analysis done in this work, we adopt the slitloss-rescaled spectra as our fiducial spectra, since matching the total photometric flux provides the most consistent estimate of integrated galaxy properties.

\subsection{Dependency on Star-Formation History Parameterization} \label{subsec:SFH_comp}

\begin{figure}[ht]
    \centering
    \includegraphics[width=\linewidth]{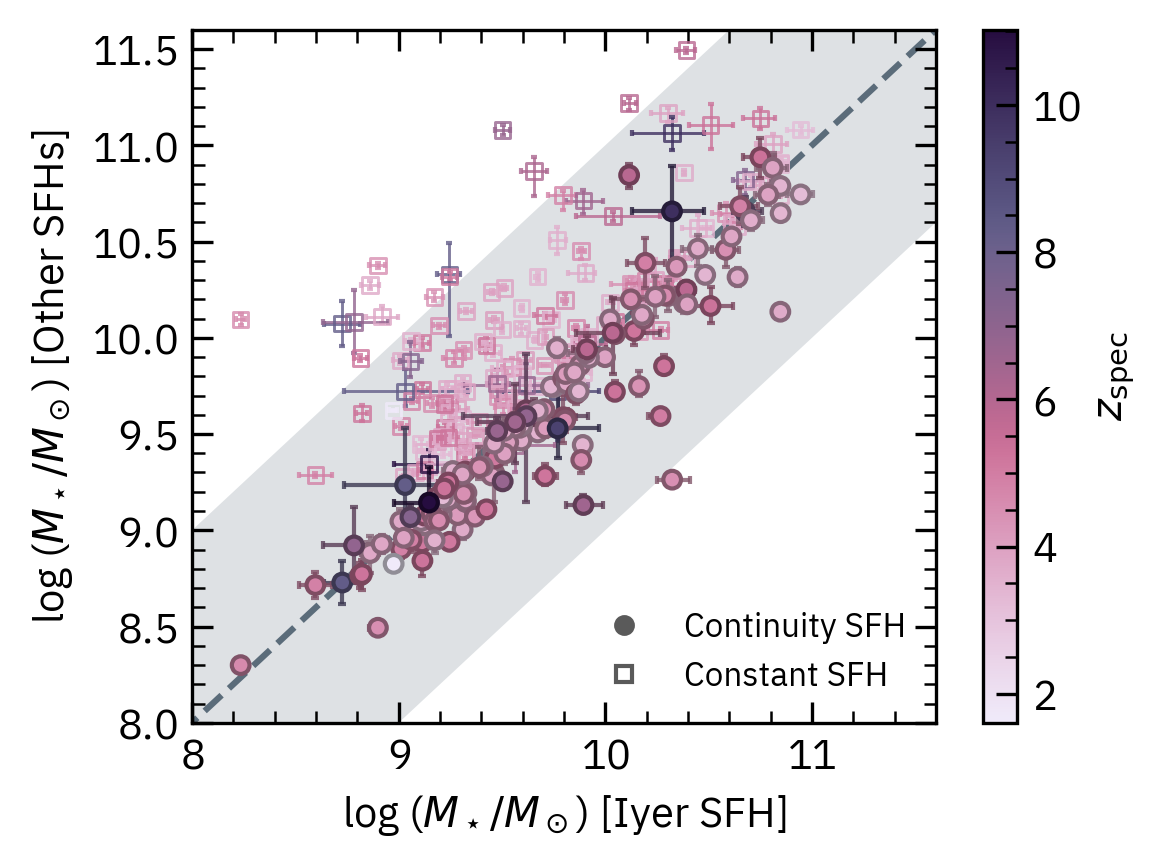}
    \caption{Comparison of stellar masses recovered using different star-formation history (SFH) parametrizations. For this work, we adopt the \citet{Iyer2019} Dense Basis SFH as the fiducial model (x-axis). Filled circles show stellar masses recovered using the \citet{Leja2019} continuity SFH, while open squares correspond to a flat (constant) SFH. The two flexible, non-parametric SFH models (Iyer and continuity) yield stellar masses in good agreement across the full mass range, indicating that the inferred stellar masses are robust to the choice between these parametrizations. In contrast, the constant SFH systematically overestimates stellar masses and fails to provide acceptable fits to the spectroscopic data.}
    \label{fig:Mstar_comp_SFH}
\end{figure}

To explore how assumptions in the parametrization of SFHs affect the physical properties inferred from SED fitting, we perform two additional fits within \bp\ aside from our fiducial model (\S~\ref{subsec:sed_fitting}). First, we adopt the \citet{Leja2019} continuity SFH with the default student-t prior, in which star formation is distributed across a set of user-defined time bins. Second, we assume a constant star-formation rate across the lifetime of the galaxy. The stellar mass recovery of different SFHs is shown in Figure~\ref{fig:Mstar_comp_SFH}. Notably, the constant SFH is unable to produce statistically acceptable fits to the observed spectra, having a $\chi^2$ of $1.4 \times$ that of the non-parametric SFHs, and we therefore exclude it as a viable SFH assumption for this sample. The continuity SFH generally performs comparably to the Dense Basis SFH, yielding similar goodness-of-fit metrics, with the median $\chi^2$ values differing by $<0.2$\%. The SFHs recovered under the two models are also broadly consistent in shape, with both reproducing the major episodes of star formation inferred for each galaxy (Figure~\ref{fig:SED_SFH}). Overall, this agreement indicates that the inferred stellar masses and SFH-dependent physical properties are robust to the choice between flexible, non-parametric SFH parametrizations, while strongly disfavoring restrictive assumptions such as a constant star-formation history.  In particular, the presence of recent star-formation bursts and the diversity of late-time mass assembly persist across flexible modeling approaches. This supports our conclusion that episodic growth and late-time assembly are not primarily driven by a specific SFH prescription, as long as the SFH is adequately flexible, but are instead required by the data. The recovered star-formation histories for all sources in this work can be found in Figure~\ref{fig:SED_SFH}, and the full figure set available in the online jounral.


\section{Discussion}  \label{sec:discussion}

\subsection{Phases of Stellar Mass Assembly}

For the galaxies presented in this work, we find that the diversity of stellar mass assembly histories decreases toward later stages of formation, with the scatter of the inferred times at which galaxies assemble 25\% of their stellar mass ($\tau_{25}$) being systematically larger than that observed at later stages ($\tau_{50}$, $\tau_{75}$). 
We do not find a strong redshift dependence in the formation timescales once the timescales are normalized to the age of the Universe. Instead, the recovered formation timescales are closely tied to sSFRs, with the galaxies with the lowest sSFRs forming the bulk of their stellar mass at earlier times. When compared to the galaxies in a $\sim 800$ arcmin$^2$ lightcone with SC-SAM galaxy formation physics \citep{Yung2022} and hydrodynamical simulation FLARES \citep{Lovell2021, Lovell2023}, we see that the observed galaxies form systematically earlier than their counterparts (matched by stellar mass and redshift at the epoch of observation) in the SC-SAM and in FLARES, however, the formation timescales converge at later galaxy epochs in both cases. 
The weak mass and redshift dependence of the median normalized assembly times suggests that the typical onset of significant star formation is largely similar across the population. However, the large scatter in $\tau_{25}$ at fixed mass indicates diversity in early growth, likely reflecting stochasticity in gas accretion or star-formation and halo assembly histories \citep[e.g.][]{Fakhouri2010, Behroozi2013,Neistein2008}. Galaxies with low sSFR completed the bulk of their stellar mass growth earlier compared to the majority of the galaxies, consistent with early gas exhaustion or feedback-driven shutdown. \citep[e.g.][]{Carnall2019, Tacchella2020, Lovell2023}. However, for the bulk of the galaxies analyzed here, scatter decreases in the later assembly stages possibly due to feedback and star formation regulating galaxy growth and suppressing early differences. 

The prevalence of recent star-formation activity inferred for the majority of the sample highlights the importance of rapid, late-time growth in building massive galaxies at high redshift. This finding reinforces emerging JWST-era results that challenge the usage of more simple, parametric SFHs commonly assumed in photometric studies, and instead favor SFHs that incorporate bursts or more flexible non-parametric prescriptions \citep[e.g.,][]{Tacchella2023, Looser2024, EstradaCarpenter2023}.

\subsection{Dust Attenuation for Massive High Redshift Galaxies}

We find that massive galaxies in our sample exhibit shallow, gray dust attenuation curves relative to the Calzetti law, with nearly all galaxies above $M_\star \sim 10^{10}\,M_\odot$ requiring attenuation curves shallower than that of the Calzetti law. Lower-mass systems span a much wider range of attenuation slopes, including steeper than SMC curves. We note that this trend is partly expected, as shallower attenuation curves permit higher $A_V$, even in UV-bright galaxies, which in turn increases the inferred stellar mass at fixed observed photometry \citep{Salim2018, Chevallard2013}.

This mass dependence closely tracks the observed correlation between attenuation slope and total dust optical depth, with galaxies at higher $A_V$ tending toward flatter attenuation curves, supported by previous studies \citep[e.g.,][]{Salim2018,Salim2020, Draine2003, Reddy2015}. The association between stellar mass, dust content, and attenuation-curve shape likely reflects a combination of increased dust column density and more complex dust–star geometries in massive systems. Radiative transfer models predict that as optical depth increases, multiple scattering and mixed dust–star geometries naturally produce grayer effective attenuation laws \citep[e.g.,][]{WittGordon2000, Chevallard2013, Narayanan2018, Sommovigo2025}. Additionally, efficient dust grain growth can occur in environments with high metallicity and gas density, biasing the grain size distribution toward larger grains and further flattening the UV attenuation \citep[e.g.,][]{Hirashita2012, Asano2013}. 

At the highest redshifts probed ($z \gtrsim 6$), the majority of the galaxies in our sample exhibit attenuation curves shallower than the SMC, and some are flatter than the Calzetti law (Figure~\ref{fig:Av_binned}). This is consistent with early dust enrichment scenarios in which core-collapse supernovae are the dominant source of dust production combined with rapid grain growth. In these scenarios, only the larger grains tend to survive the reverse shock and enter the interstellar medium, resulting in a grain size distribution skewed toward larger sizes compared to that expected from AGB enrichment \citep[e.g.,][]{Nozawa2007,Silvia2010, Hirashita2015, Witstok2023, McKinney2025, Hirashita2005}.
Variations in $\delta$ may also be attributed to environments where dust and stars are distributed in clumpy or mixed geometries \citep{WittGordon2000, CharlotFall2000, Chevallard2013}. 

While we cannot currently disentangle the relative contributions of grain physics and geometry to the grayer attenuation curves, our results do underscore the importance of adopting flexible dust models when interpreting the SEDs of massive galaxies in the early Universe. Additionally allowing for sufficiently broad priors on dust parameters is necessary to recover adequate fits. While this may inflate uncertainties on other derived quantities, it more faithfully represents the true constraints available from the data. Future theoretical work, particularly dust evolution models coupled with cosmological simulations, can help establish physically motivated priors that reduce this systematic.

\section{Summary}  \label{sec:summary}

In this work, we investigate the formation histories of massive (\logM$\gtrsim 9.5$) galaxies at $z \gtrsim 3.5$ using joint spectro-photometric modeling of \textit{JWST}/NIRSpec prism spectroscopy from the CAPERS survey combined with deep \textit{HST} and \textit{JWST} multi-wavelength imaging. The reconstruction of diverse star-formation histories in this work is enabled by the inclusion of spectroscopy. Spectroscopic constraints on the rest-frame optical continuum and nebular emission lines substantially reduce degeneracies between stellar age, dust attenuation, and star formation that are inherent to broadband photometry alone. In particular, the explicit separation of stellar continuum and emission-lines allows for better constraints in stellar mass and star-formation rate estimates, especially for strongly star-forming galaxies.

We select a sample of 148 \logM$>9.5$ (determined from photometric SED fitting) candidate massive galaxies at $z>3.5$ observed CAPERS. Their physical properties are inferred using SED fitting with \textsc{Bagpipes}, adopting the \citet{Iyer2019} non-parametric star-formation history and the \citet{Salim2018} dust attenuation model, which allows for variations in the attenuation-curve shape. We apply a wavelength-dependent slit-loss rescaling to place the spectra on a consistent absolute flux scale with the photometry, enabling robust recovery of integrated galaxy properties.

From this analysis, we find that the diversity of star-formation histories in massive galaxies is largest in the early stages of stellar mass assembly. While early assembly timescales exhibit substantial scatter, the dispersion is lower at epochs at which galaxies assemble 50\% and 75\% of their stellar mass, indicating that galaxy growth and feedback may serve a role in regulating timescales of galaxy growth. In particular, galaxies with low sSFRs have $\tau_{50}$ and $\tau_{75}$ earlier compared to the median of the galaxy sample, indicating more rapid growth. Many galaxies show evidence for recent or ongoing star-formation episodes, indicating that rapid, late-time growth  may play an important role in building stellar mass even in systems that are already massive at early cosmic times. 

We further find that dust attenuation properties vary systematically with stellar mass and dust content. Massive galaxies preferentially require shallow, gray attenuation curves relative to the SMC law, resulting a relative increase in attenuation in the UV, with the flattest curves occurring at high $A_V$ and at higher redshifts probed. This behavior is consistent with expectations from radiative transfer effects in optically thick and geometrically complex dust distributions, as well as early dust enrichment scenarios involving core-collapse supernovae and rapid grain growth in dense ISMs \citep[][e.g.]{WittGordon2000, Chevallard2013, Narayanan2018, Salim2020, Hirashita2012}. These results underscore the importance of adopting flexible dust attenuation prescriptions when interpreting the SEDs of massive galaxies in the early Universe.

Future work leveraging larger \textit{JWST} spectroscopic samples will enable more robust statistical characterization of star-formation history diversity across a wider range of stellar masses and environments. Spatially resolved IFU spectroscopy, together with constraints from far-infrared or millimeter observations, will further help disentangle the roles of dust geometry, feedback, and gas supply in regulating star formation. Together, such efforts will be essential for building a more complete picture of how galaxies assemble and quench during the first few billion years of cosmic history.

\begin{acknowledgments}
K.C. and S.L.F. acknowledge support from NASA through STScI award JWST-GO-6368. This material is based upon work supported by the National Science Foundation Graduate Research Fellowship under Grant No. DGE 2137420. 
PAH acknowledges support from NASA under award 80GSFC24M0006.
ACC acknowledges support from a UKRI Frontier Research Grantee Grant (PI Carnall; grant reference EP/Y037065/1).
This work is based on observations made with the NASA/ESA/CSA \textit{James Webb Space Telescope}, obtained at the Space Telescope Science Institute, which is operated by the Association of Universities for Research in Astronomy, Incorporated, under NASA contract NAS5-03127. Support for program number GO-6368 was provided through a grant from the STScI under NASA contract NAS5-03127. The data were obtained from the Mikulski Archive for Space Telescopes (MAST) at the Space Telescope Science Institute. 
These observations are associated with program \#6368, and can be accessed via \dataset[doi: 10.17909/0q3p-sp24].
\end{acknowledgments}

%
\facilities{\textit{HST}(STScI), \textit{JWST} (STScI}

\software{astropy \citep{2013A&A...558A..33A,2018AJ....156..123A,2022ApJ...935..167A},  
          Cloudy \citep{2013RMxAA..49..137F}, 
          Source Extractor \citep{1996A&AS..117..393B}
          }


\clearpage

\appendix

\begin{longtable}{lccccccc}
\caption{Physical properties of the CAPERS massive galaxy sample, ordered by decreasing stellar mass.}\label{tab:capers_prop}\\
\toprule
Source ID & R.A. & Decl. & $z_{\rm spec}$ & $\log(M_\star/M_\odot)$ & SFR & $A_V$ & $\delta$ \\
 & (deg) & (deg) & & (dex) & ($M_\odot\,{\rm yr}^{-1}$) & (mag) & \\
\midrule
\endfirsthead
\multicolumn{8}{c}{\tablename\ \thetable{} -- continued from previous page}\\
\toprule
Source ID & R.A. & Decl. & $z_{\rm spec}$ & $\log(M_\star/M_\odot)$ & SFR & $A_V$ & $\delta$ \\
 & (deg) & (deg) & & (dex) & ($M_\odot\,{\rm yr}^{-1}$) & (mag) & \\
\midrule
\endhead
\midrule \multicolumn{8}{r}{Continued on next page}\\
\endfoot
\bottomrule
\endlastfoot
CAPERS-UDS-21205 & 34.52105 & -5.17956 & 3.23 & $10.94^{+0.06}_{-0.07}$ & $551.63^{+43.02}_{-58.15}$ & $4.00^{+0.12}_{-0.13}$ & $0.11^{+0.03}_{-0.03}$ \\
CAPERS-EGS-12561 & 214.94782 & +52.91118 & 3.38 & $10.85^{+0.03}_{-0.03}$ & $274.48^{+21.05}_{-16.84}$ & $2.37^{+0.07}_{-0.07}$ & $0.37^{+0.02}_{-0.02}$ \\
CAPERS-COSMOS-24203 & 150.20902 & +2.34887 & 3.45 & $10.85^{+0.02}_{-0.02}$ & $242.66^{+15.50}_{-16.53}$ & $2.55^{+0.06}_{-0.05}$ & $0.29^{+0.02}_{-0.02}$ \\
CAPERS-COSMOS-14080 & 150.18372 & +2.40136 & 3.74 & $10.85^{+0.03}_{-0.03}$ & $392.07^{+42.32}_{-39.55}$ & $1.50^{+0.08}_{-0.07}$ & $0.57^{+0.02}_{-0.02}$ \\
CAPERS-COSMOS-11370 & 150.13316 & +2.41693 & 3.95 & $10.81^{+0.07}_{-0.08}$ & $296.84^{+60.44}_{-70.28}$ & $2.62^{+0.21}_{-0.20}$ & $0.01^{+0.08}_{-0.09}$ \\
CAPERS-UDS-25004 & 34.33419 & -5.19868 & 3.95 & $10.79^{+0.05}_{-0.06}$ & $324.99^{+87.71}_{-83.07}$ & $3.52^{+0.17}_{-0.18}$ & $0.25^{+0.06}_{-0.06}$ \\
CAPERS-EGS-9027 & 214.86428 & +52.87107 & 5.32 & $10.75^{+0.07}_{-0.09}$ & $195.67^{+24.76}_{-25.83}$ & $3.44^{+0.27}_{-0.28}$ & $0.15^{+0.03}_{-0.02}$ \\
CAPERS-UDS-22285 & 34.32519 & -5.18440 & 3.60 & $10.70^{+0.06}_{-0.06}$ & $263.12^{+25.55}_{-25.50}$ & $3.74^{+0.13}_{-0.15}$ & $0.23^{+0.03}_{-0.03}$ \\
CAPERS-UDS-5643 & 34.42962 & -5.11232 & 7.27 & $10.68^{+0.08}_{-0.06}$ & $403.89^{+86.07}_{-77.45}$ & $1.86^{+0.17}_{-0.15}$ & $0.14^{+0.04}_{-0.04}$ \\
CAPERS-UDS-22018 & 34.47409 & -5.18301 & 4.91 & $10.65^{+0.09}_{-0.10}$ & $119.71^{+59.31}_{-44.31}$ & $3.05^{+0.25}_{-0.23}$ & $0.13^{+0.06}_{-0.08}$ \\
CAPERS-UDS-36910 & 34.50829 & -5.25603 & 3.70 & $10.64^{+0.03}_{-0.03}$ & $221.14^{+25.02}_{-23.78}$ & $0.87^{+0.09}_{-0.09}$ & $0.17^{+0.05}_{-0.04}$ \\
CAPERS-UDS-23398 & 34.32274 & -5.19043 & 3.81 & $10.61^{+0.03}_{-0.03}$ & $158.53^{+22.06}_{-23.27}$ & $1.89^{+0.08}_{-0.09}$ & $0.38^{+0.04}_{-0.05}$ \\
CAPERS-EGS-8453 & 214.93158 & +52.92100 & 4.48 & $10.58^{+0.06}_{-0.07}$ & $130.28^{+35.06}_{-27.30}$ & $3.05^{+0.19}_{-0.17}$ & $0.31^{+0.05}_{-0.05}$ \\
CAPERS-EGS-9028 & 214.86450 & +52.87097 & 5.32 & $10.51^{+0.11}_{-0.10}$ & $263.69^{+75.95}_{-58.88}$ & $2.51^{+0.31}_{-0.24}$ & $0.40^{+0.04}_{-0.04}$ \\
CAPERS-COSMOS-19865 & 150.09019 & +2.37058 & 3.42 & $10.48^{+0.02}_{-0.02}$ & $109.38^{+6.64}_{-6.60}$ & $0.56^{+0.04}_{-0.04}$ & $-0.14^{+0.03}_{-0.04}$ \\
CAPERS-UDS-43418 & 34.31717 & -5.28553 & 3.68 & $10.45^{+0.07}_{-0.08}$ & $62.82^{+32.70}_{-23.67}$ & $3.48^{+0.18}_{-0.24}$ & $0.41^{+0.04}_{-0.06}$ \\
CAPERS-COSMOS-53606 & 150.15452 & +2.18716 & 3.96 & $10.40^{+0.02}_{-0.02}$ & $102.90^{+9.49}_{-10.71}$ & $0.83^{+0.06}_{-0.07}$ & $-0.12^{+0.04}_{-0.05}$ \\
CAPERS-COSMOS-52597 & 150.14622 & +2.19430 & 5.84 & $10.39^{+0.04}_{-0.05}$ & $262.46^{+24.50}_{-27.13}$ & $1.50^{+0.09}_{-0.08}$ & $0.25^{+0.02}_{-0.02}$ \\
CAPERS-COSMOS-4367 & 150.16562 & +2.46962 & 3.71 & $10.38^{+0.01}_{-0.01}$ & $30.08^{+3.11}_{-2.37}$ & $0.06^{+0.01}_{-0.01}$ & $-1.30^{+0.08}_{-0.06}$ \\
CAPERS-EGS-7545 & 215.03905 & +53.00278 & 4.29 & $10.34^{+0.04}_{-0.03}$ & $34.84^{+15.49}_{-11.11}$ & $0.81^{+0.08}_{-0.09}$ & $-0.49^{+0.08}_{-0.08}$ \\
CAPERS-COSMOS-42262 & 150.13582 & +2.25789 & 4.44 & $10.32^{+0.09}_{-0.08}$ & $108.79^{+14.79}_{-13.63}$ & $3.90^{+0.36}_{-0.37}$ & $0.12^{+0.01}_{-0.01}$ \\
CAPERS-EGS-25297 & 214.81711 & +52.74834 & 9.94 & $10.32^{+0.16}_{-0.19}$ & $221.33^{+97.62}_{-79.19}$ & $2.38^{+0.23}_{-0.32}$ & $0.43^{+0.01}_{-0.02}$ \\
CAPERS-UDS-6363 & 34.45242 & -5.11599 & 3.96 & $10.30^{+0.07}_{-0.08}$ & $65.27^{+28.32}_{-13.27}$ & $1.23^{+0.24}_{-0.24}$ & $0.54^{+0.05}_{-0.11}$ \\
CAPERS-EGS-21360 & 214.98134 & +52.88256 & 4.56 & $10.29^{+0.03}_{-0.03}$ & $7.57^{+2.37}_{-1.44}$ & $1.36^{+0.06}_{-0.06}$ & $0.15^{+0.04}_{-0.05}$ \\
CAPERS-EGS-12737 & 214.87820 & +52.86092 & 5.28 & $10.28^{+0.03}_{-0.03}$ & $84.52^{+16.28}_{-13.13}$ & $1.56^{+0.11}_{-0.12}$ & $-0.00^{+0.05}_{-0.05}$ \\
CAPERS-EGS-24591 & 214.94948 & +52.84541 & 5.01 & $10.27^{+0.03}_{-0.06}$ & $37.22^{+16.25}_{-4.60}$ & $1.75^{+0.12}_{-0.14}$ & $-0.05^{+0.01}_{-0.01}$ \\
CAPERS-UDS-27904 & 34.45837 & -5.21396 & 3.97 & $10.24^{+0.10}_{-0.10}$ & $36.87^{+22.86}_{-12.28}$ & $3.38^{+0.26}_{-0.26}$ & $0.04^{+0.10}_{-0.11}$ \\
CAPERS-COSMOS-9836 & 150.11319 & +2.42811 & 4.92 & $10.19^{+0.10}_{-0.09}$ & $18.46^{+16.72}_{-6.20}$ & $3.46^{+0.35}_{-0.36}$ & $0.20^{+0.09}_{-0.05}$ \\
CAPERS-UDS-19769 & 34.50344 & -5.17372 & 4.55 & $10.19^{+0.03}_{-0.04}$ & $58.04^{+11.58}_{-12.83}$ & $0.57^{+0.10}_{-0.14}$ & $0.54^{+0.04}_{-0.06}$ \\
CAPERS-EGS-21559 & 214.81329 & +52.76219 & 3.79 & $10.18^{+0.01}_{-0.01}$ & $59.37^{+7.22}_{-7.12}$ & $1.23^{+0.06}_{-0.06}$ & $0.08^{+0.03}_{-0.03}$ \\
CAPERS-UDS-37970 & 34.33162 & -5.26059 & 4.37 & $10.16^{+0.02}_{-0.02}$ & $68.63^{+6.69}_{-5.96}$ & $2.01^{+0.13}_{-0.14}$ & $-0.02^{+0.01}_{-0.01}$ \\
CAPERS-COSMOS-10452 & 150.10768 & +2.42347 & 5.24 & $10.14^{+0.05}_{-0.04}$ & $74.32^{+10.11}_{-8.92}$ & $1.25^{+0.13}_{-0.15}$ & $0.19^{+0.05}_{-0.04}$ \\
CAPERS-UDS-33421 & 34.44432 & -5.24039 & 4.43 & $10.12^{+0.08}_{-0.10}$ & $93.47^{+17.72}_{-18.43}$ & $2.89^{+0.20}_{-0.22}$ & $0.15^{+0.04}_{-0.04}$ \\
CAPERS-COSMOS-51267 & 150.12382 & +2.20337 & 5.89 & $10.11^{+0.03}_{-0.03}$ & $139.06^{+11.48}_{-8.50}$ & $1.72^{+0.19}_{-0.13}$ & $0.39^{+0.01}_{-0.01}$ \\
CAPERS-COSMOS-10639 & 150.11310 & +2.42209 & 4.66 & $10.05^{+0.02}_{-0.02}$ & $39.50^{+4.37}_{-5.13}$ & $0.59^{+0.09}_{-0.12}$ & $-0.19^{+0.03}_{-0.05}$ \\
CAPERS-EGS-7061 & 214.98571 & +52.96756 & 5.29 & $10.05^{+0.03}_{-0.03}$ & $64.36^{+10.80}_{-6.98}$ & $1.43^{+0.14}_{-0.16}$ & $0.13^{+0.03}_{-0.04}$ \\
CAPERS-COSMOS-42364 & 150.10977 & +2.25737 & 5.86 & $10.04^{+0.23}_{-0.18}$ & $109.15^{+64.48}_{-37.09}$ & $2.86^{+0.63}_{-0.40}$ & $0.23^{+0.07}_{-0.07}$ \\
CAPERS-UDS-35916 & 34.51197 & -5.25155 & 3.71 & $10.02^{+0.02}_{-0.01}$ & $40.77^{+3.32}_{-2.92}$ & $0.69^{+0.02}_{-0.03}$ & $-1.01^{+0.07}_{-0.07}$ \\
CAPERS-UDS-20999 & 34.32529 & -5.17875 & 3.95 & $10.00^{+0.01}_{-0.01}$ & $33.66^{+7.03}_{-5.94}$ & $0.59^{+0.05}_{-0.06}$ & $-0.28^{+0.05}_{-0.05}$ \\
CAPERS-UDS-28669 & 34.31962 & -5.21815 & 3.55 & $9.92^{+0.03}_{-0.03}$ & $14.81^{+2.19}_{-1.59}$ & $0.15^{+0.06}_{-0.04}$ & $-0.80^{+0.17}_{-0.18}$ \\
CAPERS-COSMOS-50793 & 150.11511 & +2.20649 & 5.84 & $9.91^{+0.09}_{-0.07}$ & $18.48^{+13.86}_{-5.27}$ & $0.80^{+0.19}_{-0.15}$ & $-0.38^{+0.15}_{-0.20}$ \\
CAPERS-UDS-36732 & 34.50995 & -5.25515 & 3.71 & $9.90^{+0.04}_{-0.08}$ & $17.51^{+2.51}_{-1.19}$ & $1.77^{+0.20}_{-0.34}$ & $0.09^{+0.05}_{-0.05}$ \\
CAPERS-EGS-1034 & 214.86907 & +52.91889 & 6.53 & $9.89^{+0.09}_{-0.09}$ & $35.27^{+5.75}_{-5.53}$ & $2.05^{+0.29}_{-0.30}$ & $0.05^{+0.02}_{-0.03}$ \\
CAPERS-EGS-3024 & 214.99735 & +52.99917 & 3.47 & $9.89^{+0.04}_{-0.04}$ & $37.42^{+7.18}_{-6.03}$ & $1.03^{+0.10}_{-0.09}$ & $0.46^{+0.04}_{-0.04}$ \\
CAPERS-UDS-5510 & 34.43569 & -5.11181 & 4.55 & $9.88^{+0.02}_{-0.02}$ & $10.94^{+1.38}_{-0.96}$ & $0.05^{+0.01}_{-0.01}$ & $-1.32^{+0.08}_{-0.06}$ \\
CAPERS-UDS-23982 & 34.48151 & -5.19343 & 4.09 & $9.88^{+0.02}_{-0.01}$ & $15.93^{+1.61}_{-1.12}$ & $0.11^{+0.04}_{-0.02}$ & $-0.86^{+0.19}_{-0.16}$ \\
CAPERS-UDS-32921 & 34.42154 & -5.23817 & 3.40 & $9.87^{+0.02}_{-0.02}$ & $19.38^{+1.54}_{-1.24}$ & $0.16^{+0.03}_{-0.03}$ & $-0.72^{+0.10}_{-0.09}$ \\
CAPERS-UDS-29620 & 34.46713 & -5.22254 & 4.81 & $9.86^{+0.01}_{-0.01}$ & $16.79^{+1.29}_{-1.03}$ & $0.06^{+0.03}_{-0.01}$ & $-1.13^{+0.24}_{-0.17}$ \\
CAPERS-UDS-40230 & 34.32100 & -5.27056 & 3.80 & $9.85^{+0.04}_{-0.04}$ & $25.23^{+7.11}_{-8.96}$ & $0.74^{+0.12}_{-0.14}$ & $0.27^{+0.11}_{-0.11}$ \\
CAPERS-COSMOS-13301 & 150.13256 & +2.40577 & 3.75 & $9.81^{+0.02}_{-0.02}$ & $23.33^{+3.00}_{-4.37}$ & $0.18^{+0.06}_{-0.07}$ & $-0.25^{+0.17}_{-0.22}$ \\
CAPERS-COSMOS-35830 & 150.17749 & +2.29146 & 5.11 & $9.81^{+0.01}_{-0.01}$ & $13.24^{+0.90}_{-1.10}$ & $0.07^{+0.03}_{-0.02}$ & $-1.04^{+0.17}_{-0.21}$ \\
CAPERS-COSMOS-37601 & 150.16420 & +2.28243 & 4.09 & $9.81^{+0.01}_{-0.01}$ & $16.91^{+1.69}_{-1.70}$ & $0.34^{+0.06}_{-0.06}$ & $-0.44^{+0.06}_{-0.08}$ \\
CAPERS-COSMOS-37844 & 150.13687 & +2.28123 & 4.60 & $9.80^{+0.01}_{-0.01}$ & $7.30^{+1.30}_{-0.69}$ & $0.05^{+0.01}_{-0.01}$ & $-1.30^{+0.10}_{-0.07}$ \\
CAPERS-EGS-1170 & 214.93287 & +52.96321 & 4.90 & $9.80^{+0.06}_{-0.07}$ & $34.04^{+3.69}_{-4.16}$ & $1.92^{+0.15}_{-0.17}$ & $0.02^{+0.03}_{-0.02}$ \\
CAPERS-EGS-13283 & 214.94942 & +52.90787 & 4.57 & $9.80^{+0.02}_{-0.02}$ & $14.86^{+1.97}_{-1.42}$ & $0.05^{+0.01}_{-0.01}$ & $-1.26^{+0.14}_{-0.09}$ \\
CAPERS-UDS-60476 & 34.52097 & -5.29268 & 9.30 & $9.77^{+0.20}_{-0.27}$ & $16.58^{+19.66}_{-10.13}$ & $1.42^{+0.54}_{-0.67}$ & $0.48^{+0.04}_{-0.10}$ \\
CAPERS-UDS-22286 & 34.32518 & -5.18451 & 3.59 & $9.77^{+0.04}_{-0.03}$ & $42.05^{+3.92}_{-3.12}$ & $2.68^{+0.15}_{-0.09}$ & $0.31^{+0.04}_{-0.04}$ \\
CAPERS-EGS-8485 & 214.92249 & +52.91532 & 2.94 & $9.75^{+0.05}_{-0.05}$ & $2.73^{+0.51}_{-1.39}$ & $2.98^{+0.18}_{-0.16}$ & $0.13^{+0.03}_{-0.02}$ \\
CAPERS-EGS-11174 & 214.86483 & +52.86050 & 3.43 & $9.74^{+0.02}_{-0.01}$ & $15.81^{+1.89}_{-1.34}$ & $0.06^{+0.03}_{-0.01}$ & $-1.10^{+0.25}_{-0.20}$ \\
CAPERS-UDS-20283 & 34.29117 & -5.17565 & 4.14 & $9.71^{+0.01}_{-0.01}$ & $4.04^{+2.19}_{-0.60}$ & $0.27^{+0.04}_{-0.04}$ & $-0.74^{+0.08}_{-0.07}$ \\
CAPERS-UDS-22365 & 34.33891 & -5.18479 & 5.24 & $9.71^{+0.06}_{-0.05}$ & $29.01^{+5.20}_{-4.75}$ & $1.33^{+0.23}_{-0.20}$ & $0.14^{+0.05}_{-0.06}$ \\
CAPERS-EGS-11968 & 214.85882 & +52.85147 & 4.36 & $9.70^{+0.02}_{-0.02}$ & $15.00^{+1.34}_{-1.21}$ & $0.09^{+0.03}_{-0.02}$ & $-0.79^{+0.16}_{-0.18}$ \\
CAPERS-UDS-36012 & 34.46689 & -5.25194 & 3.55 & $9.67^{+0.03}_{-0.02}$ & $18.22^{+1.27}_{-1.20}$ & $0.64^{+0.05}_{-0.05}$ & $-0.35^{+0.05}_{-0.05}$ \\
CAPERS-UDS-11655 & 34.42451 & -5.13923 & 3.50 & $9.67^{+0.01}_{-0.01}$ & $7.62^{+4.97}_{-2.27}$ & $0.06^{+0.01}_{-0.02}$ & $-1.03^{+0.15}_{-0.18}$ \\
CAPERS-EGS-12067 & 214.86633 & +52.85628 & 3.64 & $9.66^{+0.01}_{-0.01}$ & $4.75^{+4.00}_{-1.45}$ & $0.06^{+0.01}_{-0.01}$ & $-1.36^{+0.05}_{-0.03}$ \\
CAPERS-EGS-19394 & 214.95841 & +52.87511 & 6.16 & $9.65^{+0.06}_{-0.06}$ & $36.63^{+7.48}_{-5.68}$ & $1.18^{+0.18}_{-0.15}$ & $0.45^{+0.02}_{-0.03}$ \\
CAPERS-COSMOS-38491 & 150.10644 & +2.27803 & 6.86 & $9.62^{+0.30}_{-0.30}$ & $9.26^{+8.41}_{-3.64}$ & $2.10^{+0.67}_{-0.64}$ & $0.26^{+0.13}_{-0.11}$ \\
CAPERS-UDS-24166 & 34.46705 & -5.19442 & 6.12 & $9.62^{+0.03}_{-0.03}$ & $8.05^{+2.29}_{-1.27}$ & $0.11^{+0.07}_{-0.04}$ & $-0.19^{+0.21}_{-0.27}$ \\
CAPERS-UDS-2064 & 34.22146 & -5.09431 & 4.16 & $9.61^{+0.03}_{-0.03}$ & $13.23^{+0.67}_{-0.62}$ & $0.05^{+0.01}_{-0.01}$ & $-1.23^{+0.16}_{-0.11}$ \\
CAPERS-UDS-29648 & 34.32640 & -5.22272 & 3.32 & $9.60^{+0.02}_{-0.02}$ & $10.84^{+1.59}_{-1.25}$ & $0.08^{+0.02}_{-0.02}$ & $-1.10^{+0.15}_{-0.14}$ \\
CAPERS-UDS-12122 & 34.47655 & -5.14131 & 3.55 & $9.59^{+0.01}_{-0.01}$ & $13.27^{+0.62}_{-0.94}$ & $0.04^{+0.01}_{-0.01}$ & $-1.32^{+0.09}_{-0.06}$ \\
CAPERS-EGS-16317 & 214.93277 & +52.87610 & 3.65 & $9.59^{+0.02}_{-0.02}$ & $12.70^{+1.49}_{-0.82}$ & $0.03^{+0.01}_{-0.01}$ & $-1.11^{+0.21}_{-0.18}$ \\
CAPERS-UDS-25034 & 34.50601 & -5.19874 & 3.70 & $9.59^{+0.01}_{-0.01}$ & $12.29^{+1.57}_{-1.95}$ & $0.33^{+0.14}_{-0.13}$ & $-0.34^{+0.10}_{-0.21}$ \\
CAPERS-COSMOS-11739 & 150.14272 & +2.41544 & 4.68 & $9.57^{+0.02}_{-0.02}$ & $6.49^{+1.10}_{-0.62}$ & $0.12^{+0.06}_{-0.03}$ & $-0.80^{+0.22}_{-0.17}$ \\
CAPERS-UDS-38694 & 34.32982 & -5.26404 & 3.80 & $9.56^{+0.02}_{-0.02}$ & $9.39^{+1.14}_{-1.17}$ & $0.30^{+0.07}_{-0.08}$ & $-0.22^{+0.09}_{-0.11}$ \\
CAPERS-UDS-98865 & 34.31377 & -5.20575 & 6.47 & $9.56^{+0.19}_{-0.18}$ & $8.30^{+10.05}_{-4.26}$ & $1.26^{+0.44}_{-0.51}$ & $0.36^{+0.14}_{-0.25}$ \\
CAPERS-UDS-4854 & 34.43742 & -5.10882 & 4.55 & $9.55^{+0.04}_{-0.05}$ & $13.67^{+2.09}_{-1.42}$ & $0.94^{+0.11}_{-0.10}$ & $0.01^{+0.12}_{-0.09}$ \\
CAPERS-COSMOS-28222 & 150.15495 & +2.32858 & 4.05 & $9.53^{+0.02}_{-0.03}$ & $10.26^{+1.21}_{-1.10}$ & $0.04^{+0.01}_{-0.01}$ & $-1.26^{+0.14}_{-0.09}$ \\
CAPERS-COSMOS-29019 & 150.12706 & +2.32483 & 3.97 & $9.51^{+0.01}_{-0.02}$ & $8.23^{+1.95}_{-1.85}$ & $0.05^{+0.01}_{-0.01}$ & $-1.34^{+0.06}_{-0.04}$ \\
CAPERS-COSMOS-39352 & 150.09824 & +2.27336 & 3.68 & $9.51^{+0.06}_{-0.06}$ & $13.89^{+2.36}_{-2.32}$ & $1.11^{+0.14}_{-0.14}$ & $0.06^{+0.09}_{-0.10}$ \\
CAPERS-EGS-25077 & 214.94918 & +52.84318 & 6.73 & $9.50^{+0.03}_{-0.04}$ & $14.85^{+0.87}_{-0.90}$ & $0.51^{+0.07}_{-0.07}$ & $-0.29^{+0.05}_{-0.06}$ \\
CAPERS-COSMOS-14863 & 150.07636 & +2.39728 & 4.68 & $9.50^{+0.04}_{-0.04}$ & $12.73^{+3.67}_{-3.36}$ & $0.06^{+0.03}_{-0.02}$ & $-1.03^{+0.33}_{-0.22}$ \\
CAPERS-EGS-11894 & 214.86553 & +52.85663 & 3.64 & $9.50^{+0.01}_{-0.01}$ & $2.53^{+1.12}_{-0.29}$ & $0.03^{+0.01}_{-0.01}$ & $-1.31^{+0.10}_{-0.07}$ \\
CAPERS-EGS-18152 & 214.86403 & +52.81536 & 4.39 & $9.49^{+0.01}_{-0.01}$ & $20.41^{+1.21}_{-1.51}$ & $0.04^{+0.01}_{-0.01}$ & $-1.33^{+0.09}_{-0.05}$ \\
CAPERS-EGS-3760 & 214.87241 & +52.90632 & 4.52 & $9.48^{+0.03}_{-0.03}$ & $9.39^{+1.06}_{-1.00}$ & $0.19^{+0.06}_{-0.04}$ & $-0.42^{+0.14}_{-0.17}$ \\
CAPERS-COSMOS-50585 & 150.10325 & +2.20798 & 6.69 & $9.48^{+0.05}_{-0.05}$ & $12.55^{+4.40}_{-2.08}$ & $0.08^{+0.05}_{-0.04}$ & $-0.22^{+0.23}_{-0.27}$ \\
CAPERS-COSMOS-26929 & 150.14622 & +2.33566 & 5.09 & $9.47^{+0.02}_{-0.02}$ & $10.43^{+2.02}_{-2.04}$ & $0.37^{+0.12}_{-0.13}$ & $-0.30^{+0.11}_{-0.18}$ \\
CAPERS-UDS-45025 & 34.48550 & -5.29359 & 3.72 & $9.46^{+0.02}_{-0.02}$ & $6.47^{+0.51}_{-0.79}$ & $0.07^{+0.02}_{-0.01}$ & $-1.24^{+0.15}_{-0.11}$ \\
CAPERS-COSMOS-8000 & 150.13879 & +2.44144 & 4.15 & $9.46^{+0.02}_{-0.02}$ & $13.05^{+1.58}_{-1.47}$ & $0.68^{+0.07}_{-0.07}$ & $-0.13^{+0.03}_{-0.02}$ \\
CAPERS-UDS-16602 & 34.50471 & -5.15954 & 3.60 & $9.46^{+0.03}_{-0.03}$ & $14.44^{+1.81}_{-1.70}$ & $0.84^{+0.20}_{-0.23}$ & $-0.11^{+0.07}_{-0.07}$ \\
CAPERS-COSMOS-9005 & 150.16979 & +2.43388 & 3.99 & $9.45^{+0.01}_{-0.02}$ & $5.09^{+1.01}_{-0.42}$ & $0.05^{+0.01}_{-0.01}$ & $-1.34^{+0.06}_{-0.04}$ \\
CAPERS-EGS-28516 & 214.86692 & +52.76928 & 5.29 & $9.42^{+0.02}_{-0.02}$ & $12.22^{+1.04}_{-0.80}$ & $0.59^{+0.07}_{-0.06}$ & $-0.29^{+0.03}_{-0.03}$ \\
CAPERS-UDS-34858 & 34.46384 & -5.24681 & 3.65 & $9.41^{+0.02}_{-0.02}$ & $8.04^{+0.37}_{-0.39}$ & $0.05^{+0.01}_{-0.01}$ & $-1.24^{+0.15}_{-0.10}$ \\
CAPERS-COSMOS-38539 & 150.10966 & +2.27753 & 4.39 & $9.39^{+0.08}_{-0.05}$ & $14.13^{+2.04}_{-3.42}$ & $0.08^{+0.09}_{-0.05}$ & $-0.10^{+0.35}_{-0.41}$ \\
CAPERS-EGS-28257 & 214.94853 & +52.82726 & 4.27 & $9.37^{+0.02}_{-0.02}$ & $13.44^{+1.45}_{-1.07}$ & $0.03^{+0.01}_{-0.01}$ & $-1.22^{+0.22}_{-0.13}$ \\
CAPERS-COSMOS-27439 & 150.16069 & +2.33291 & 3.97 & $9.33^{+0.02}_{-0.02}$ & $9.55^{+1.40}_{-1.11}$ & $0.10^{+0.03}_{-0.03}$ & $-0.91^{+0.16}_{-0.16}$ \\
CAPERS-EGS-24654 & 214.81959 & +52.75271 & 3.92 & $9.32^{+0.01}_{-0.01}$ & $5.66^{+0.55}_{-0.56}$ & $0.03^{+0.01}_{-0.01}$ & $-1.31^{+0.09}_{-0.06}$ \\
CAPERS-COSMOS-16464 & 150.08995 & +2.38846 & 3.93 & $9.32^{+0.03}_{-0.03}$ & $9.83^{+0.68}_{-0.46}$ & $0.04^{+0.01}_{-0.01}$ & $-1.03^{+0.18}_{-0.19}$ \\
CAPERS-EGS-25315 & 214.87544 & +52.78981 & 5.28 & $9.32^{+0.02}_{-0.03}$ & $4.79^{+0.83}_{-0.54}$ & $0.12^{+0.08}_{-0.05}$ & $-0.41^{+0.22}_{-0.28}$ \\
CAPERS-COSMOS-33076 & 150.11941 & +2.30483 & 4.44 & $9.31^{+0.02}_{-0.02}$ & $8.53^{+0.36}_{-0.26}$ & $0.03^{+0.01}_{-0.01}$ & $-1.22^{+0.24}_{-0.13}$ \\
CAPERS-COSMOS-7950 & 150.16336 & +2.44190 & 3.37 & $9.31^{+0.03}_{-0.02}$ & $11.69^{+0.66}_{-0.91}$ & $0.03^{+0.03}_{-0.01}$ & $-0.15^{+0.53}_{-0.72}$ \\
CAPERS-COSMOS-9441 & 150.17763 & +2.43080 & 3.73 & $9.31^{+0.02}_{-0.02}$ & $4.99^{+1.26}_{-0.99}$ & $0.09^{+0.01}_{-0.01}$ & $-1.31^{+0.08}_{-0.06}$ \\
CAPERS-UDS-36375 & 34.45612 & -5.25365 & 3.85 & $9.31^{+0.02}_{-0.02}$ & $4.98^{+0.81}_{-0.51}$ & $0.04^{+0.01}_{-0.01}$ & $-1.29^{+0.10}_{-0.08}$ \\
CAPERS-UDS-10442 & 34.26782 & -5.13457 & 4.11 & $9.29^{+0.05}_{-0.05}$ & $10.19^{+0.64}_{-0.85}$ & $0.08^{+0.04}_{-0.03}$ & $-0.94^{+0.24}_{-0.22}$ \\
CAPERS-EGS-8700 & 214.92648 & +52.91681 & 3.60 & $9.27^{+0.03}_{-0.03}$ & $6.31^{+0.99}_{-0.46}$ & $0.03^{+0.01}_{-0.01}$ & $-1.16^{+0.23}_{-0.16}$ \\
CAPERS-EGS-14944 & 214.88802 & +52.85327 & 4.83 & $9.27^{+0.05}_{-0.06}$ & $6.94^{+1.73}_{-0.89}$ & $0.10^{+0.11}_{-0.06}$ & $0.26^{+0.24}_{-0.36}$ \\
CAPERS-UDS-4831 & 34.25131 & -5.10872 & 3.51 & $9.26^{+0.05}_{-0.05}$ & $9.08^{+0.69}_{-0.69}$ & $0.05^{+0.02}_{-0.01}$ & $-1.10^{+0.18}_{-0.17}$ \\
CAPERS-UDS-32520 & 34.42827 & -5.23617 & 5.14 & $9.25^{+0.03}_{-0.04}$ & $9.05^{+1.48}_{-1.41}$ & $0.04^{+0.01}_{-0.01}$ & $-1.30^{+0.14}_{-0.08}$ \\
CAPERS-UDS-13189 & 34.41929 & -5.14598 & 8.28 & $9.25^{+0.05}_{-0.06}$ & $7.55^{+1.84}_{-0.77}$ & $0.06^{+0.01}_{-0.01}$ & $-1.27^{+0.16}_{-0.09}$ \\
CAPERS-UDS-16685 & 34.29313 & -5.15990 & 4.71 & $9.24^{+0.04}_{-0.04}$ & $5.28^{+0.69}_{-0.41}$ & $0.06^{+0.05}_{-0.02}$ & $-0.53^{+0.32}_{-0.34}$ \\
CAPERS-UDS-25000 & 34.51677 & -5.19862 & 3.73 & $9.24^{+0.03}_{-0.03}$ & $4.99^{+0.53}_{-0.29}$ & $0.02^{+0.02}_{-0.01}$ & $-0.88^{+0.32}_{-0.29}$ \\
CAPERS-UDS-41008 & 34.47074 & -5.27383 & 5.20 & $9.24^{+0.05}_{-0.05}$ & $7.56^{+2.61}_{-2.08}$ & $0.09^{+0.07}_{-0.05}$ & $0.33^{+0.19}_{-0.27}$ \\
CAPERS-EGS-19842 & 215.02418 & +52.91969 & 3.74 & $9.23^{+0.03}_{-0.02}$ & $5.77^{+0.34}_{-0.33}$ & $0.10^{+0.02}_{-0.02}$ & $-0.80^{+0.15}_{-0.13}$ \\
CAPERS-UDS-10729 & 34.21739 & -5.13531 & 5.33 & $9.22^{+0.05}_{-0.03}$ & $4.97^{+1.42}_{-0.54}$ & $0.05^{+0.07}_{-0.02}$ & $-0.24^{+0.51}_{-0.66}$ \\
CAPERS-COSMOS-45305 & 150.15169 & +2.24004 & 5.30 & $9.22^{+0.04}_{-0.03}$ & $7.45^{+0.91}_{-0.81}$ & $0.34^{+0.08}_{-0.07}$ & $-0.09^{+0.11}_{-0.11}$ \\
CAPERS-COSMOS-39158 & 150.14149 & +2.27442 & 3.24 & $9.21^{+0.03}_{-0.04}$ & $4.01^{+1.26}_{-0.75}$ & $0.07^{+0.06}_{-0.04}$ & $0.13^{+0.34}_{-0.37}$ \\
CAPERS-UDS-22960 & 34.46015 & -5.18792 & 3.81 & $9.21^{+0.03}_{-0.04}$ & $3.88^{+0.71}_{-0.42}$ & $0.16^{+0.05}_{-0.04}$ & $-0.53^{+0.18}_{-0.19}$ \\
CAPERS-UDS-15420 & 34.47251 & -5.15465 & 3.61 & $9.21^{+0.03}_{-0.02}$ & $7.16^{+0.33}_{-0.32}$ & $0.07^{+0.02}_{-0.01}$ & $-1.25^{+0.15}_{-0.11}$ \\
CAPERS-UDS-41375 & 34.49102 & -5.27537 & 4.56 & $9.19^{+0.03}_{-0.03}$ & $9.53^{+0.31}_{-0.58}$ & $0.04^{+0.02}_{-0.01}$ & $-1.15^{+0.16}_{-0.17}$ \\
CAPERS-UDS-22752 & 34.48497 & -5.18678 & 3.87 & $9.19^{+0.03}_{-0.03}$ & $5.07^{+0.68}_{-0.33}$ & $0.02^{+0.01}_{-0.01}$ & $-1.09^{+0.21}_{-0.19}$ \\
CAPERS-COSMOS-36884 & 150.13612 & +2.28580 & 5.35 & $9.19^{+0.03}_{-0.03}$ & $3.94^{+0.26}_{-0.19}$ & $0.04^{+0.02}_{-0.01}$ & $-1.19^{+0.25}_{-0.14}$ \\
CAPERS-COSMOS-45724 & 150.06912 & +2.23764 & 3.54 & $9.17^{+0.04}_{-0.04}$ & $6.19^{+1.45}_{-1.47}$ & $0.06^{+0.05}_{-0.03}$ & $-0.29^{+0.49}_{-0.56}$ \\
CAPERS-COSMOS-39408 & 150.10428 & +2.27302 & 4.38 & $9.17^{+0.04}_{-0.04}$ & $8.19^{+0.33}_{-0.56}$ & $0.03^{+0.01}_{-0.01}$ & $-1.26^{+0.19}_{-0.10}$ \\
CAPERS-UDS-35695 & 34.30895 & -5.25060 & 4.63 & $9.16^{+0.02}_{-0.02}$ & $5.60^{+0.27}_{-0.21}$ & $0.03^{+0.02}_{-0.01}$ & $-1.18^{+0.22}_{-0.16}$ \\
CAPERS-UDS-126973 & 34.26444 & -5.09623 & 11.00 & $9.15^{+0.16}_{-0.17}$ & $11.36^{+5.64}_{-4.00}$ & $0.06^{+0.06}_{-0.03}$ & $-1.05^{+0.47}_{-0.22}$ \\
CAPERS-EGS-22256 & 214.94841 & +52.85510 & 5.25 & $9.12^{+0.03}_{-0.02}$ & $10.36^{+0.71}_{-0.83}$ & $0.04^{+0.05}_{-0.02}$ & $0.30^{+0.23}_{-0.37}$ \\
CAPERS-EGS-14848 & 214.80285 & +52.79315 & 3.51 & $9.12^{+0.03}_{-0.02}$ & $5.10^{+0.63}_{-0.52}$ & $0.03^{+0.04}_{-0.02}$ & $-0.24^{+0.41}_{-0.44}$ \\
CAPERS-EGS-12328 & 214.81304 & +52.81698 & 5.28 & $9.11^{+0.03}_{-0.03}$ & $6.58^{+0.30}_{-0.34}$ & $0.08^{+0.03}_{-0.02}$ & $-1.01^{+0.20}_{-0.17}$ \\
CAPERS-EGS-6009 & 215.01329 & +52.99325 & 5.07 & $9.11^{+0.04}_{-0.04}$ & $7.16^{+0.28}_{-0.48}$ & $0.03^{+0.01}_{-0.01}$ & $-1.19^{+0.22}_{-0.14}$ \\
CAPERS-COSMOS-21526 & 150.08992 & +2.36244 & 3.37 & $9.11^{+0.03}_{-0.03}$ & $3.89^{+0.93}_{-0.45}$ & $0.02^{+0.01}_{-0.01}$ & $-1.05^{+0.59}_{-0.24}$ \\
CAPERS-UDS-19208 & 34.30301 & -5.17125 & 3.88 & $9.08^{+0.02}_{-0.02}$ & $5.67^{+0.77}_{-0.49}$ & $0.05^{+0.01}_{-0.01}$ & $-1.15^{+0.21}_{-0.16}$ \\
CAPERS-COSMOS-34068 & 150.17223 & +2.30004 & 4.91 & $9.06^{+0.03}_{-0.03}$ & $4.43^{+0.20}_{-0.21}$ & $0.03^{+0.01}_{-0.01}$ & $-1.18^{+0.33}_{-0.14}$ \\
CAPERS-UDS-7353 & 34.45780 & -5.12071 & 3.60 & $9.06^{+0.02}_{-0.03}$ & $5.93^{+0.22}_{-0.31}$ & $0.04^{+0.01}_{-0.01}$ & $-1.31^{+0.12}_{-0.06}$ \\
CAPERS-UDS-95782 & 34.42813 & -5.21394 & 7.25 & $9.06^{+0.05}_{-0.05}$ & $5.77^{+0.76}_{-0.70}$ & $0.18^{+0.07}_{-0.04}$ & $-0.72^{+0.22}_{-0.15}$ \\
CAPERS-UDS-95796 & 34.32275 & -5.21391 & 8.28 & $9.03^{+0.30}_{-0.29}$ & $5.10^{+2.44}_{-2.10}$ & $1.18^{+0.49}_{-0.43}$ & $0.18^{+0.11}_{-0.13}$ \\
CAPERS-COSMOS-36295 & 150.14026 & +2.28843 & 3.93 & $9.02^{+0.02}_{-0.01}$ & $4.96^{+0.47}_{-0.41}$ & $0.03^{+0.01}_{-0.01}$ & $-1.31^{+0.11}_{-0.06}$ \\
CAPERS-EGS-401 & 214.93485 & +52.96908 & 4.89 & $9.01^{+0.03}_{-0.03}$ & $4.59^{+0.21}_{-0.21}$ & $0.04^{+0.01}_{-0.01}$ & $-1.16^{+0.19}_{-0.16}$ \\
CAPERS-EGS-15850 & 214.88061 & +52.84211 & 3.76 & $9.01^{+0.03}_{-0.03}$ & $4.81^{+0.26}_{-0.37}$ & $0.05^{+0.04}_{-0.02}$ & $-0.61^{+0.36}_{-0.39}$ \\
CAPERS-COSMOS-40855 & 150.13540 & +2.26538 & 1.60 & $8.98^{+0.02}_{-0.02}$ & $4.03^{+0.31}_{-0.24}$ & $0.18^{+0.02}_{-0.02}$ & $-1.36^{+0.05}_{-0.03}$ \\
CAPERS-COSMOS-26481 & 150.15405 & +2.33808 & 3.83 & $8.92^{+0.08}_{-0.08}$ & $5.40^{+0.46}_{-0.58}$ & $0.05^{+0.03}_{-0.02}$ & $-0.72^{+0.32}_{-0.29}$ \\
CAPERS-UDS-27017 & 34.32229 & -5.20921 & 4.50 & $8.90^{+0.04}_{-0.04}$ & $3.33^{+0.33}_{-0.25}$ & $0.17^{+0.10}_{-0.08}$ & $-0.24^{+0.12}_{-0.16}$ \\
CAPERS-COSMOS-16820 & 150.12271 & +2.38648 & 3.76 & $8.86^{+0.04}_{-0.05}$ & $4.94^{+0.51}_{-0.58}$ & $0.05^{+0.03}_{-0.02}$ & $-0.77^{+0.35}_{-0.30}$ \\
CAPERS-COSMOS-14738 & 150.12492 & +2.39804 & 5.25 & $8.82^{+0.03}_{-0.03}$ & $3.66^{+0.26}_{-0.26}$ & $0.09^{+0.03}_{-0.03}$ & $-0.90^{+0.22}_{-0.25}$ \\
CAPERS-EGS-14297 & 214.80092 & +52.79549 & 5.29 & $8.81^{+0.03}_{-0.03}$ & $4.24^{+0.29}_{-0.63}$ & $0.02^{+0.02}_{-0.01}$ & $-0.97^{+0.52}_{-0.28}$ \\
CAPERS-COSMOS-38872 & 150.10357 & +2.27607 & 6.87 & $8.78^{+0.16}_{-0.15}$ & $4.26^{+1.79}_{-1.19}$ & $0.57^{+0.29}_{-0.22}$ & $0.01^{+0.18}_{-0.19}$ \\
CAPERS-EGS-2145 & 214.95194 & +52.97174 & 8.31 & $8.73^{+0.09}_{-0.09}$ & $3.50^{+0.85}_{-0.65}$ & $0.34^{+0.16}_{-0.09}$ & $-0.29^{+0.16}_{-0.17}$ \\
CAPERS-EGS-17203 & 214.83143 & +52.79830 & 4.88 & $8.60^{+0.08}_{-0.08}$ & $2.99^{+0.38}_{-0.41}$ & $0.07^{+0.02}_{-0.01}$ & $-0.97^{+0.13}_{-0.13}$ \\
CAPERS-EGS-7401 & 215.02608 & +52.99451 & 4.63 & $8.24^{+0.02}_{-0.02}$ & $1.83^{+0.10}_{-0.09}$ & $0.07^{+0.02}_{-0.01}$ & $-0.84^{+0.18}_{-0.15}$ \\
\end{longtable}

\clearpage

\bibliography{paper}{}
\bibliographystyle{aasjournalv7}



\end{document}